\documentclass{aa}    

\input epsf
\usepackage{longtable,graphicx} 

\begin{document}

\def\HW{HW\,CMa}
\def\SW{SW\,CMa}
\def\kms{\ifmmode{\rm km\thinspace s^{-1}}\else km\thinspace s$^{-1}$\fi}
\def\phs{\phantom{$-$}}

\title{Absolute dimensions of eclipsing binaries. XXIX.  \thanks{Based
on observations carried out with the Nordic Optical Telescope (NOT) at
La Palma, the 50\,cm Str\"omgren Automatic Telescope (SAT) at ESO, La
Silla, the 1.5\,m Wyeth reflector at the Oak Ridge Observatory, Harvard,
Massachusetts, USA, and the 1.5\,m Tillinghast reflector at the F.\ L.\
Whipple Observatory, Mt.\ Hopkins, Arizona, USA.} }

\subtitle{The Am-type systems SW\,Canis Majoris and HW\,Canis Majoris.
\thanks{Tables~\ref{tab:swcma_rv} and \ref{tab:hwcma_rv} as well as Appendices B and C will be
available in electronic form at the CDS via anonymous ftp to
130.79.128.5 or via http://cdsweb.u-strasbg.fr/Abstract.html.}}

\author{
G.\ Torres \inst{1}
\and J.\ V.\ Clausen \inst{2}\thanks{Deceased.}
\and H.\ Bruntt \inst{2,3}
\and A.\ Claret \inst{4}
\and J.\ Andersen \inst{2}
\and B.\ Nordstr\"om \inst{2}
\and R.\ P.\ Stefanik \inst{1}
\and D.~W.\ Latham \inst{1}
}
\offprints{G. Torres, \\ e-mail: gtorres@cfa.harvard.edu}

\institute{
Harvard-Smithsonian Center for Astrophysics, 60 Garden Street,
Cambridge, MA 02138, USA
\and
Niels Bohr Institute, Copenhagen University, Juliane Maries Vej 30,
DK-2100 Copenhagen {\O}, Denmark
\and
School of Physics A28, University of Sydney, 2006 NSW, Australia
\and
Instituto de Astrof\'\i sica de Andaluc\'\i a, Apartado 3004, 18080
Granada, Spain
}

\date{Received 02 Aug 2011 / Accepted xx Dec 2011}
 
\titlerunning{\SW\ and \HW}
\authorrunning{G.\ Torres et al.}

\abstract
{Accurate physical properties of eclipsing stars provide important
constraints on models of stellar structure and evolution, especially
when combined with spectroscopic information on their chemical
composition. Empirical calibrations of the data also lead to accurate
mass and radius estimates for exoplanet host stars. Finally, accurate
data for unusual stellar subtypes, such as Am stars, also help to
unravel the cause(s) of their peculiarities.}
{We aim to determine the masses, radii, effective temperatures,
detailed chemical composition and rotational speeds for the Am-type
eclipsing binaries \SW\ (A4-5m) and \HW\ (A6m) and compare them with
similar normal stars.}
{Accurate radial velocities from the Digital Speedometers of the
Harvard-Smithsonian Center for Astrophysics were combined with
previously published $uvby$ photometry to determine precise physical
parameters for the four stars.  A detailed abundance analysis was
performed from high-resolution spectra obtained with the Nordic
Optical Telescope (La Palma).}
{We find the masses of the (relatively evolved) stars in \SW\ to be
2.10 and 2.24\,$M_{\sun}$, with radii of 2.50 and 3.01\,$R_{\sun}$,
while the (essentially zero-age) stars in \HW\ have masses of 1.72 and
1.78\,$M_{\sun}$, radii of 1.64 and 1.66\,$R_{\sun}$ -- all with
errors well below 2\%. Detailed atmospheric abundances for one or both
components were determined for 14 elements in \SW\ ([Fe/H] =
+0.49/+0.61 dex) and 16 in \HW\ ([Fe/H] = +0.33/+0.32 dex); both
abundance patterns are characteristic of metallic-line stars. Both
systems are well fit by current stellar evolution models for assumed
bulk abundances of [Fe/H] = +0.05 and +0.23, respectively
([$\alpha$/Fe] = 0.0), and ages of $\sim$700~Myr and 160~Myr.}
{}


\maketitle


\section{Introduction}
\label{sec:intro}

Among the eclipsing binaries with the best determined fundamental
properties approximately two dozen are of spectral type A (see Torres
et al.\ \cite{Torres:10a}).  Roughly 20--30\% of main-sequence A stars
are known to be metallic-lined (Am). The Am stars are found
overwhelmingly to be in binary systems (e.g., Abt \cite{Abt:61}). They
are characterized by peculiar abundance patterns including a
deficiency of light elements such as Sc and Ca, enhancement of
iron-peak elements, and a strong overabundance of Sr and some rare
earths, sometimes by an order of magnitude or more. The enhancement
generally increases with increasing atomic number. Their typically
slower axial rotation and stronger metal lines compared to normal A
stars tend to facilitate the radial-velocity measurements necessary
for accurate mass determinations.  It is therefore not surprising that
systems with Am stars are over-represented among the best-measured
eclipsing binaries: fully half of the A-type systems for which we know
the absolute masses and radii with relative errors less than 3\%
contain at least one component showing these abundance abnormalities.

Although significant progress in understanding and modeling these
chemically peculiar objects has been made in recent years, important
questions regarding the role of rotation and the origin of the
detailed patterns of element enhancement and depletion are not yet
completely settled (see, e.g., Abt \cite{Abt:00}, B\"ohm-Vitense
\cite{Bohm-Vitense:06}, Vick et al.\ \cite{Vick:10}). Further
observational constraints are highly desirable, particularly ones in
which very accurate mass and radius determinations are accompanied by
a detailed abundance study of the components.

\begin{figure}[b]
\resizebox{\hsize}{!}{\includegraphics{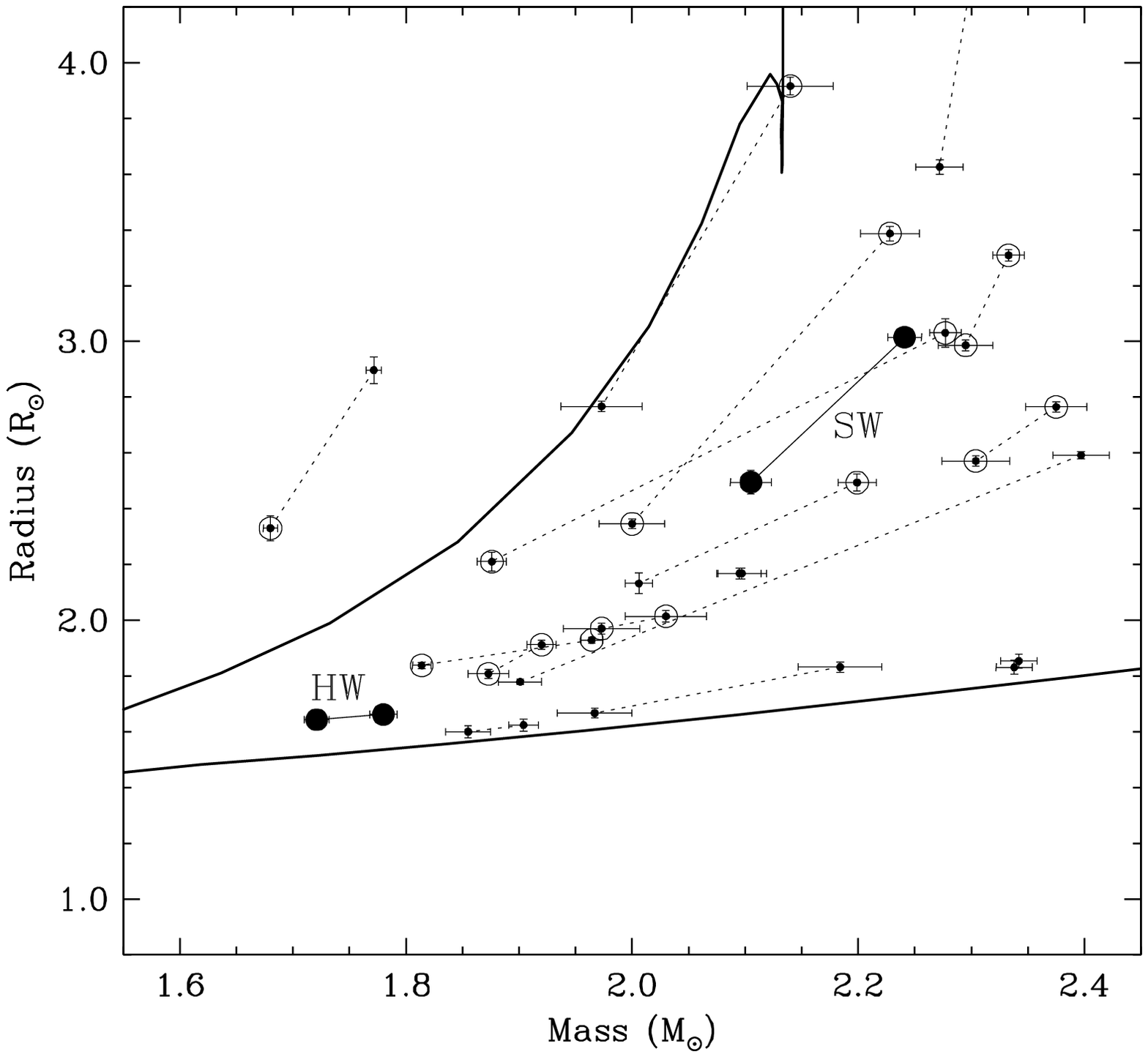}}
\epsfxsize=80mm
\caption{Main sequence eclipsing binaries taken from Torres et al.\
(\cite{Torres:10a}) in which both components have masses in the range
1.6--2.4 $M_{\sun}$, and relative errors in the masses and radii of
3\% or better. Primary and secondary components are connected with
dotted lines, and stars that are confirmed or probable Am stars are
marked with open circles.  \SW\ and \HW\ are shown with filled
symbols.  The thick solid lines are drawn for reference, and
correspond to solar-metallicity isochrones from the series of Yi et
al.\ (\cite{Yi:01}) for ages of 50~Myr and 1~Gyr.}
\label{fig:debs}
\end{figure}

Beyond a qualitative assessment of their metallic-lined nature,
individual element abundances are known for precious few of the dozen
or so Am binaries with well determined properties (see, e.g.,
Lyubimkov et al.\ \cite{Lyubimkov:96}). Thus, little can be said about
whether or how the detailed abundance patterns that seem to vary to
some degree from star to star depend on global properties such as
mass, effective temperature, or surface gravity. In this paper we
present extensive new observations of two eclipsing Am systems, \SW\
and \HW, with the goal of determining not only accurate absolute
dimensions but also individual abundances for all four stars. It is
hoped that these results and others like them will present an
opportunity for advancement on some of the issues mentioned above.
Figure~\ref{fig:debs} places the two systems in the context of all
other main sequence eclipsing binaries in the same mass regime that
have the best-determined masses and radii. Am stars are marked with
open circles. \SW\ is seen to be among the more evolved cases, while
\HW\ is the least evolved in this mass range.

We describe below our spectroscopic observations of these systems, and
an analysis of those data along with previously-published
high-precision four-color Str\"omgren photometry in order to obtain
accurate masses and radii. For \SW\ our determinations substantially
improve upon previous estimates; for \HW\ they represent the first
such measurements. We conclude with a detailed comparison with models
of stellar and tidal evolution.


\section{SW\,CMa}
\label{sec:swcma}

The observational history of \SW\ (HD\,54520, HIP\,34431, $V = 9.15$,
A4--A5) has been summarized by Clausen et al.\ (\cite{Clausen:08}).
Since its discovery as an eclipsing binary by Hoffmeister
(\cite{Hoffmeister:32}), relatively little progress in understanding
its physical properties was made in more than six decades until the
most recent study by Lacy (\cite{Lacy:97}). In that work photoelectric
$UBV$ light curves were combined with new spectroscopy to yield the
first precise determination of the masses and radii for the
components. The orbit is eccentric, and apsidal motion was detected by
Lacy (\cite{Lacy:97}) and subsequently refined by Clausen et al.\
(\cite{Clausen:08}), although the apsidal period is very long and
still poorly constrained ($U = 14\,900 \pm 4700$ yr).

For the analysis in this paper we adopt the linear ephemeris derived
by Clausen et al.\ (\cite{Clausen:08}),
\begin{displaymath}
\label{eq:swcma_eph}
{\rm Min~I~(HJD)} = 2\,446\,829.6482(1) + 10\fd091988(5) \times E~,
\end{displaymath}
in which the uncertainties in the period and reference epoch are
indicated in parentheses in units of the last decimal place. This
corresponds to the eclipse of the more massive and slightly hotter
component. The secondary eclipse occurs approximately at phase 0.31.


\subsection{Radial velocities and spectroscopic orbit}
\label{sec:swcma_spec_orb}

The spectroscopic observations of \SW\ used here for the
radial-velocity determinations were carried out at the
Harvard-Smithsonian Center for Astrophysics (CfA) between 1988 April
and 1989 March. A total of 39 spectra were gathered with the 1.5\,m
Wyeth reflector at the Oak Ridge Observatory (Harvard, Massachusetts).
Two additional spectra were taken using the 1.5\,m Tillinghast
reflector at the F.\ L.\ Whipple Observatory on Mt.\ Hopkins,
Arizona. Nearly identical spectrographs (``Digital Speedometers''
(DS); Latham \cite{Latham:92}) were used on each telescope, equipped
with photon-counting Reticon detectors that recorded a single echelle
order 45\,\AA\ wide centered at a wavelength near 5187\,\AA\
(\ion{Mg}{i}~b triplet). The resolving power of these instruments is
$\lambda/\Delta\lambda \approx 35\,000$, and the signal-to-noise (S/N)
ratios of the 41 spectra we collected range from 10 to 25 per
resolution element of 8.5~\kms.

Radial velocities were obtained with TODCOR, a two-dimensional
cross-correlation technique introduced by Zucker \& Mazeh
(\cite{Zucker:94}). Templates for the cross-correlations were selected
from a library of synthetic spectra based on model atmospheres by R.\
L.\ Kurucz, which has been described by Nordstr\"om et al.\
(\cite{Nordstrom:94}) and Latham et al.\ (\cite{Latham:02}). The
parameters of these templates are the effective temperature ($T_{\rm
eff}$), metallicity ([m/H]), rotational velocity ($v \sin i$, when
seen in projection), and surface gravity ($\log g$). 

The optimal templates for the two components were determined by
running extensive grids of cross-correlations and selecting the
combination of parameters yielding the highest correlation averaged
over all exposures, weighted by the strength of each spectrum (see
Torres et al.\ \cite{Torres:02}). The surface gravities were held
fixed at $\log g = 4.0$ for both stars, close to the final values in
Sect.~\ref{sec:swcma_absdim}. Temperatures and rotational velocities
were optimized for a range of different (fixed) metallicities, and the
first hint that the stars have anomalously strong metal lines came
from the fact that the highest cross-correlation value was obtained
for the largest metallicity available in our grid, [m/H] = +0.5, for
both components.  It is possible that a better match to the observed
spectra could be found for an even higher metallicity, and from the
trend of the improvement with [m/H], the effect appears somewhat more
pronounced for the secondary.

By interpolation the optimal temperatures were found to be 8210\,K and
7860\,K, and the best $v \sin i$ values are 29~\kms\ and
15~\kms. However, other photometric and spectroscopic estimates
discussed later point toward very similar $T_{\rm eff}$ values for the
two stars, as opposed to the 350\,K difference implied above.  Our
primary star $T_{\rm eff}$ estimate is in agreement with those
determinations, but the secondary value seems too cool.
Due to the limited wavelength coverage of the CfA spectra, metallicity
is strongly correlated with effective temperature so that an almost
equally good fit to the spectra can be found for different template
combinations in which the stronger lines corresponding to a metal-rich
composition can be made to appear weaker by increasing the
temperature, and vice versa. Therefore, even though we cannot go any
higher in [m/H] in our optimization scheme, a satisfactory match to
the observed spectra can still be found by adjusting the
temperature. This is likely the explanation for the excessively cool
$T_{\rm eff}$ we derive for the secondary from these spectra, which
supports the notion that this star may have a more anomalous
(generally more metal-rich) composition than the primary.
Experience shows that this degeneracy has little impact on the radial
velocities themselves, although it does of course limit our ability to
simultaneously infer accurate values of [m/H] and $T_{\rm eff}$.  For
the velocity determinations we adopted [m/H] = +0.5 for both stars,
along with temperatures of 8250\,K and 7750\,K and $v \sin i$
parameters of 30~\kms\ and 16~\kms\ for the primary and secondary.
These are the values in our grid nearest to the optimal estimates
mentioned earlier. We do not attach any astrophysical meaning to these
$T_{\rm eff}$ values, but treat them merely as adjustable parameters
intended to provide the best radial velocities, as free as possible
from biases.

\begin{figure}[t]
\resizebox{\hsize}{!}{\includegraphics{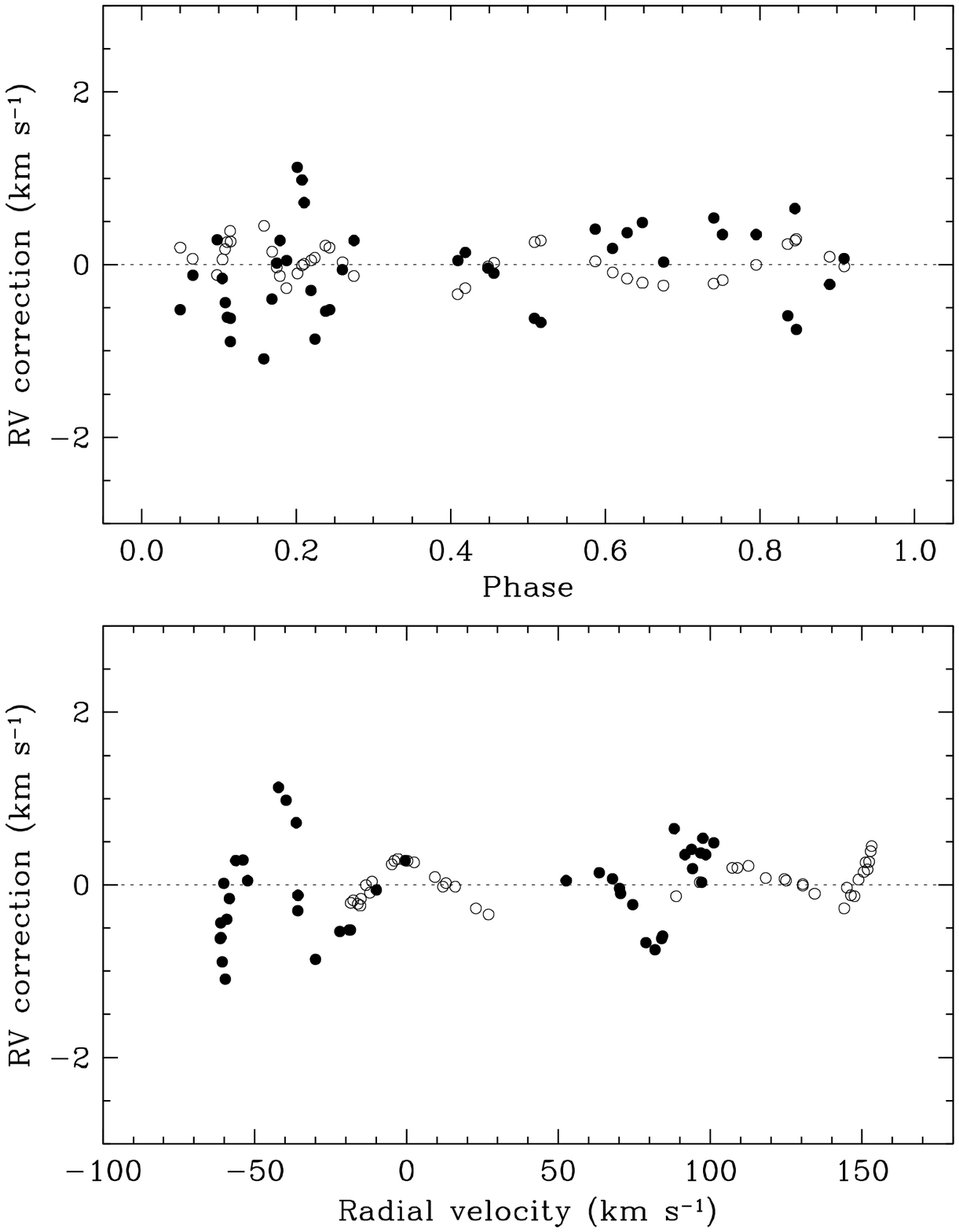}}
\epsfxsize=80mm
\caption{Systematic errors to the raw TODCOR velocities of \SW\
determined from simulations based on synthetic binary spectra. Filled
circles correspond to the primary, and open circles to the secondary.
The velocity differences are shown as a function of orbital phase and
radial velocity, and have been applied to the raw velocities as
corrections. Phase 0.0 corresponds to the primary eclipse.}
\label{fig:swcorr}
\end{figure}

The stability of the velocity zero point during the period of
observation was monitored by taking exposures of the dusk and dawn
sky. Small run-to-run corrections were applied in the manner described
by Latham (\cite{Latham:92}). Additional systematic errors may occur
because of the narrow wavelength coverage of the CfA spectra and the
fact that lines move in and out of the spectral window as a function
of orbital phase and the heliocentric correction (see Latham et al.\
{\cite{Latham:96}). To investigate this we performed experiments with
simulated double-lined spectra following the procedure detailed by
Torres et al.\ (\cite{Torres:97}), and corrections based on them were
then applied to the raw velocities. For \SW\ these corrections are
typically small ($< 1$~\kms), and their net effect is to increase the
absolute masses by slightly less than 0.4\%, and the radii by about
0.1\%. They are shown in Fig.~\ref{fig:swcorr}, where the systematic
pattern is obvious, particularly in the lower panel.

\begin{table}

\caption{Spectroscopic orbital solution for \SW.}
\label{tab:swcma_orb} 
\begin{center}
\begin{tabular}{lr}
\hline\hline\noalign{\smallskip}
\multicolumn{1}{c}{Parameter} &
\multicolumn{1}{c}{Value} \\
\noalign{\smallskip}\hline\noalign{\smallskip}
\multicolumn{2}{l}{Adjusted quantities:}  \\
~~~~$K_p$ (\kms)             & $ 82.09 \pm 0.35 $  \\
~~~~$K_s$ (\kms)             & $ 87.37 \pm 0.22 $  \\
~~~~$\gamma$ (\kms)          & $+42.57 \pm 0.17 $  \\
~~~~$e$                      & $ 0.3174 \pm 0.0011 $ \\
~~~~$\omega$ $(\degr)$ $^{\mathrm{a}}$      & $ 163.26 \pm 0.51 $ \\
\noalign{\smallskip}
\multicolumn{2}{l}{Adopted quantities:}  \\
~~~~$P$ (days)                      & 10.091988 \\
~~~~$T_{\rm I}$ (HJD$-$2\,400\,000) $^{\mathrm{b}}$   &  46829.6482 \\
\noalign{\smallskip}
\multicolumn{2}{l}{Derived quantities:}  \\
~~~~$M_p \sin^3i$ ($M_{\sun}$)      & $ 2.237  \pm 0.014  $ \\
~~~~$M_s \sin^3i$ ($M_{\sun}$)      & $ 2.102  \pm 0.018  $ \\
~~~~$q \equiv M_s/M_p $             & $ 0.9396 \pm 0.0048 $ \\
~~~~$a_p \sin i$ ($10^6$~km)        & $10.803  \pm 0.046  $ \\
~~~~$a_s \sin i$ ($10^6$~km)        & $11.498  \pm 0.029  $ \\
~~~~$a \sin i$ ($R_{\sun}$)         & $32.057   \pm 0.076  $ \\
\noalign{\smallskip}
\multicolumn{2}{l}{Other quantities pertaining to the fit:}  \\
~~~~$N_{obs}$                      &    41 \\
~~~~Time span (days)               &   333 \\
~~~~$\sigma_p$ (\kms)              &  1.93 \\
~~~~$\sigma_s$ (\kms)              &  1.19 \\
\noalign{\smallskip}
\hline
\end{tabular}
\begin{list}{}{}
\item[$^{\mathrm{a}}$] Longitude of periastron for the more massive star.
\item[$^{\mathrm{b}}$] Time of central primary eclipse (eclipse of the more massive star).
\end{list}
\end{center}
\end{table}

The final radial velocities referred to the heliocentric frame are
presented in Table~\ref{tab:swcma_rv} of Appendix~\ref{ap:rvtabs}
(available electronically), and include all corrections mentioned
above. The elements of the orbital solution we obtain from them are
listed in Table~\ref{tab:swcma_orb}, along with derived quantities
including the minimum masses and semimajor axes. The measurements and
the fitted orbit are shown in Fig.~\ref{fig:sworbit} together with the
residuals. The larger rms residual for the primary, despite being the
brighter star, is explained by the significantly higher rotational
broadening compared to the secondary. We note that while our secondary
velocity semi-amplitude agrees with the value reported by Lacy
(\cite{Lacy:97}), our primary semi-amplitude is slightly larger (by
2\%, a 1.6-$\sigma$ difference).

\begin{figure}[t]
\resizebox{\hsize}{!}{\includegraphics{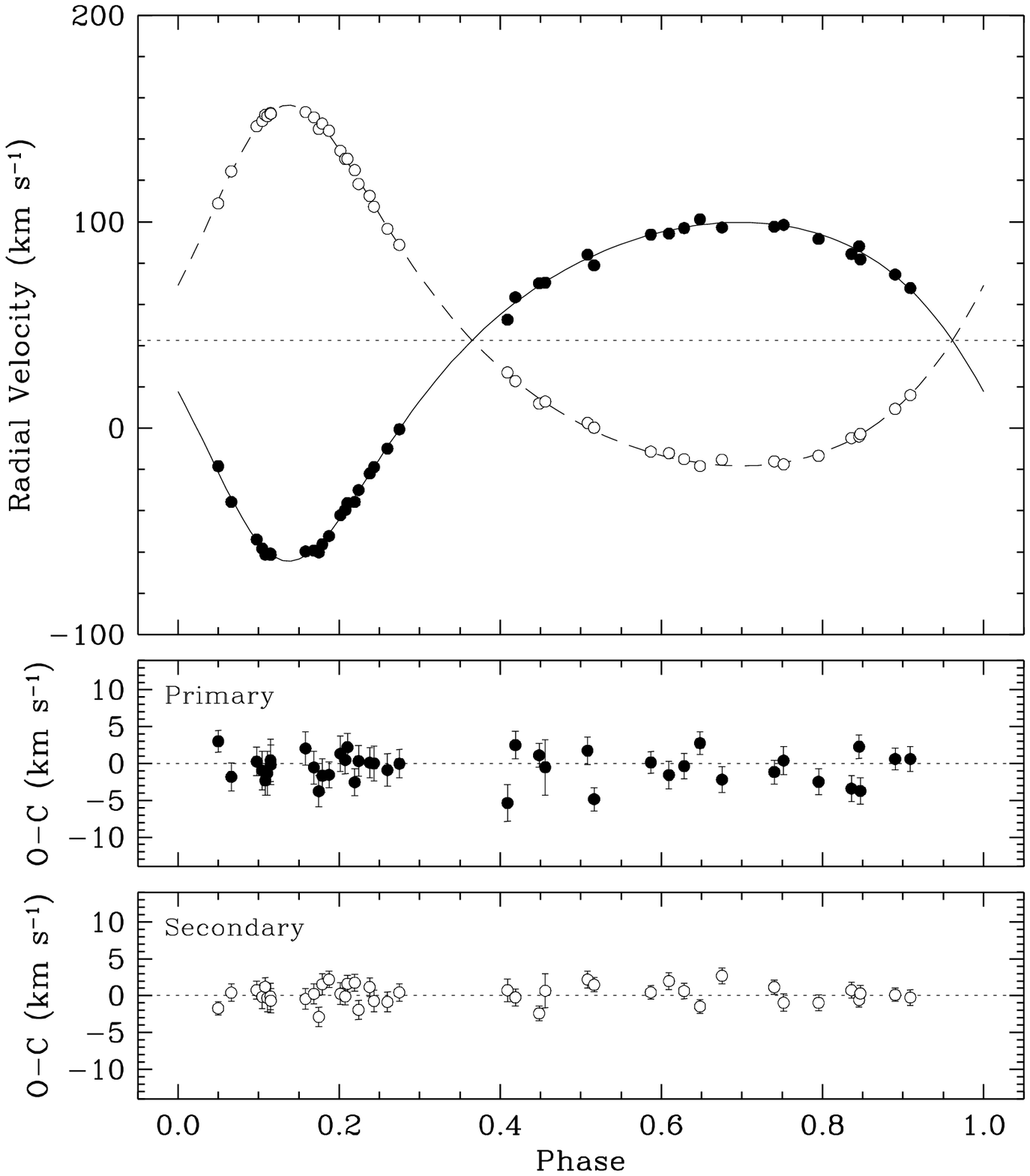}}
\epsfxsize=80mm
\caption{Measured radial velocities for \SW\ and orbital solution
(solid line for the primary, dashed for the secondary). The dotted
line in the top panel indicates the center-of-mass velocity of the
binary, and phase 0.0 corresponds to the primary eclipse. Residuals
are shown at the bottom.}
\label{fig:sworbit}
\end{figure}

In addition to the radial velocities, we used TODCOR to derive the
light ratio at the mean wavelength of our observations (5187\,\AA),
following Zucker \& Mazeh (\cite{Zucker:94}). We obtain $L_s/L_p
= 0.68 \pm 0.04$ (secondary/primary).


\subsection{Chemical abundances}
\label{sec:swcma_abund}

In order to obtain reliable estimates of the spectroscopic properties,
a spectrum of \SW\ was obtained on UT 2008 November 17 (mid exposure
at HJD 2\,454\,787.65730, orbital phase 0.5472) using the FIES
instrument (Frandsen \& Lindberg \cite{Frandsen:99}) on the 2.5\,m
Nordic Optical Telescope (NOT) on La Palma, with a resolving power of
$\lambda/\Delta\lambda \approx 46\,000$. The signal-to-noise ratio
achieved is about 60 per pixel in the 6160\,\AA\ region, and the
wavelength coverage is approximately 3600--7400\,\AA\ recorded in 78
echelle orders. Reductions used the IRAF-based FIEStool
package{\footnote{See {\tt http://www.not.iac.es} for details on FIES
and FIEStool.} with nightly flatfield, bias, and dark frames, as well
as thorium-argon exposures taken immediately before and after the
science exposure.

The spectroscopic analysis was carried out with the IDL-based VWA
analysis tool, extended to the analysis of double-lined spectra (for
details of the procedure, we refer the reader to Bruntt et al.\
\cite{Bruntt:04}, \cite{Bruntt:08}, \cite{Bruntt:09}). Briefly, VWA
uses the SYNTH software (Valenti \& Piskunov \cite{vp96}) to generate
synthetic spectra. Atmosphere models are interpolated from the recent
grid of MARCS model atmospheres (Gustafsson et al.\
\cite{Gustafsson:08}), which adopt the solar composition by Grevesse
et al.\ (\cite{Grevesse:07}).  Line information was taken from the
Vienna Atomic Line Database (VALD; Kupka et al.\ \cite{kupka99}).  In
deriving abundances relative to the Sun, the $\log gf$ values were
adjusted in such a way that each measured line in the Wallace et al.\
(\cite{whl98}) solar atlas reproduces the atmospheric abundances by
Grevesse et al.\ (\cite{Grevesse:07}).

The number of unblended lines for the analysis is somewhat limited in
this case due to the relatively large rotational broadening. We
restricted the measurements to lines with equivalent widths smaller
than about 65\,m\AA\ in the composite spectrum (corresponding to about
110\,m\AA\ for the primary and 160\,m\AA\ for the secondary), and the
surface gravities were held fixed to preliminary values near those
given in later Table~\ref{tab:swcma_absdim}. We assumed a light ratio
of $L_s/L_p = 0.69$ ($V$ band), from the light-curve solutions
described below. The effective temperatures and microturbulent
velocities ($v_{\rm mic} = 2.70$~\kms\ and 2.85~\kms\ for the primary
and secondary, with estimated errors of 0.8~\kms) were adjusted to
yield consistent values for the abundances of \ion{Fe}{i} and
\ion{Fe}{ii}, and to avoid correlations with the equivalent width and
excitation potential. The results for $T_{\rm eff}$ indicate similar
values for the primary and secondary of $8200 \pm 150$\,K and $8100
\pm 150$\,K, respectively, corresponding approximately to spectral
types A4 or A5.

\begin{table}
\caption{Abundances for the components of \SW.}
\label{tab:swcma_abundances}
\begin{center}
\begin{tabular}{lcccccc}
\hline\hline\noalign{\smallskip}
& \multicolumn{3}{c}{Primary} & \multicolumn{3}{c}{Secondary} \\
\noalign{\smallskip}\hline\noalign{\smallskip}
Element & [X/H] & $\sigma$ & $N$ & [X/H] & $\sigma$ & $N$ \\
\noalign{\smallskip}\hline\noalign{\smallskip}
\ion{Na}{i}  &   +0.57 &        &   1 &          &        &      \\
\ion{Mg}{i}  &   +0.21 &        &   1 &          &        &      \\
\ion{Si}{i}  &         &        &     &  $-$0.51 &        &   2  \\
\ion{Si}{ii} &   +0.20 &        &   2 &          &        &      \\
\ion{Ca}{i}  &   +0.05 & 0.17   &   3 &  $-$0.20 & 0.16   &   4  \\
\ion{Ti}{ii} &   +0.02 & 0.21   &   3 &    +0.37 & 0.12   &   3  \\
\ion{Cr}{ii} &   +0.15 &        &   2 &    +0.19 &        &   1  \\
\ion{Fe}{i}  &   +0.50 & 0.09   &  17 &    +0.59 & 0.10   &  22  \\
\ion{Fe}{ii} &   +0.47 & 0.09   &  11 &    +0.63 & 0.11   &  15  \\
\ion{Ni}{i}  &   +0.51 &        &   2 &    +0.42 &        &   2  \\
\ion{Zn}{i}  &   +1.08 &        &   2 &    +1.06 &        &   1  \\
\ion{Sr}{ii} &   +1.00 &        &   1 &    +1.29 &        &   1  \\
\ion{Y}{ii}  &   +1.02 &        &   1 &    +1.11 &        &   1  \\
\ion{Ba}{ii} &   +1.40 & 0.21   &   3 &    +1.54 &        &   2  \\
\noalign{\smallskip}\hline
\end{tabular}
\end{center}

\textsc{Note:} These abundances are based on $v_{\rm mic} =
2.70$~\kms\ and 2.85~\kms\ for the primary and secondary,
respectively, and [Fe/H] = +0.45 for the model atmospheres applied in
the analysis. \ion{Fe}{i} abundances include corrections for NLTE. $N$
represents the number of lines of each element, and $\sigma$ is the
scatter of the measurements (listed only if $N > 2$).
\end{table}

Individual abundances on the scale of the Grevesse et al.\
(\cite{Grevesse:07}) solar abundances were determined separately in
one or both stars for 14 species. These results are listed in
Table~\ref{tab:swcma_abundances}, along with the number of spectral
lines used in each case. Abundances based on fewer than three lines
are somewhat more uncertain. NLTE effects become significant in stars
with temperatures hotter than about 8000\,K, so appropriate
corrections to the abundances of \ion{Fe}{i} (+0.13 dex and +0.11 dex
for the primary and secondary) have been included in the values we
report, following Rentzsch-Holm (\cite{Rentzsch-Holm:96}). The average
iron abundances for the components are [Fe/H] $= +0.49 \pm 0.15$ and
$+0.61 \pm 0.15$, where the uncertainties include a contribution
from the errors in temperature and $v_{\rm mic}$.
Figure~\ref{fig:swcma_abundances} displays the abundance pattern for
both stars in \SW. Although some of these determinations are
rather uncertain due to the small number of lines and limited
signal-to-noise ratio of our spectroscopic material, the enhanced
iron-peak abundances compared to calcium, and especially the strong
overabundance of heavy elements such as Sr, Y, and Ba, are typical of
Am stars and suggest that both components are chemically
peculiar.  The iron abundance, as well as the Fe/Ca difference (0.44
and 0.81 dex), is somewhat larger for the secondary than the primary,
consistent with the hints we saw in the CfA spectra. The projected
rotational velocities were measured from the FIES spectrum by
synthesizing line profiles for a dozen isolated lines and comparing
them with the observed spectrum, seeking to minimize the residual
differences. The adopted macroturbulence velocity was $\zeta_{\rm RT}
= 8$\,\kms\ for both components, extrapolated from the trend with
temperature reported by Gray (\cite{Gray:05}). We obtained $v \sin i$
values of $24.0 \pm 1.5$~\kms\ and $10.0 \pm 1.0$~\kms\ for the
primary and secondary, respectively.  That these rotational
velocities are much slower than typical for A stars in the field is a
general characteristic found in other Am stars as well (see,
e.g., Abt \& Morrell \cite{Abt:95}, Abt \cite{Abt:00}, Fossati et al.\
\cite{Fossati:08}).\footnote{Note, however, that slow rotation in
an A star is not always associated with chemical peculiarity (see,
e.g., Fekel et al.\ \cite{Fekel:06}).}

\begin{figure}[t]
\resizebox{\hsize}{!}{\includegraphics[angle=0]{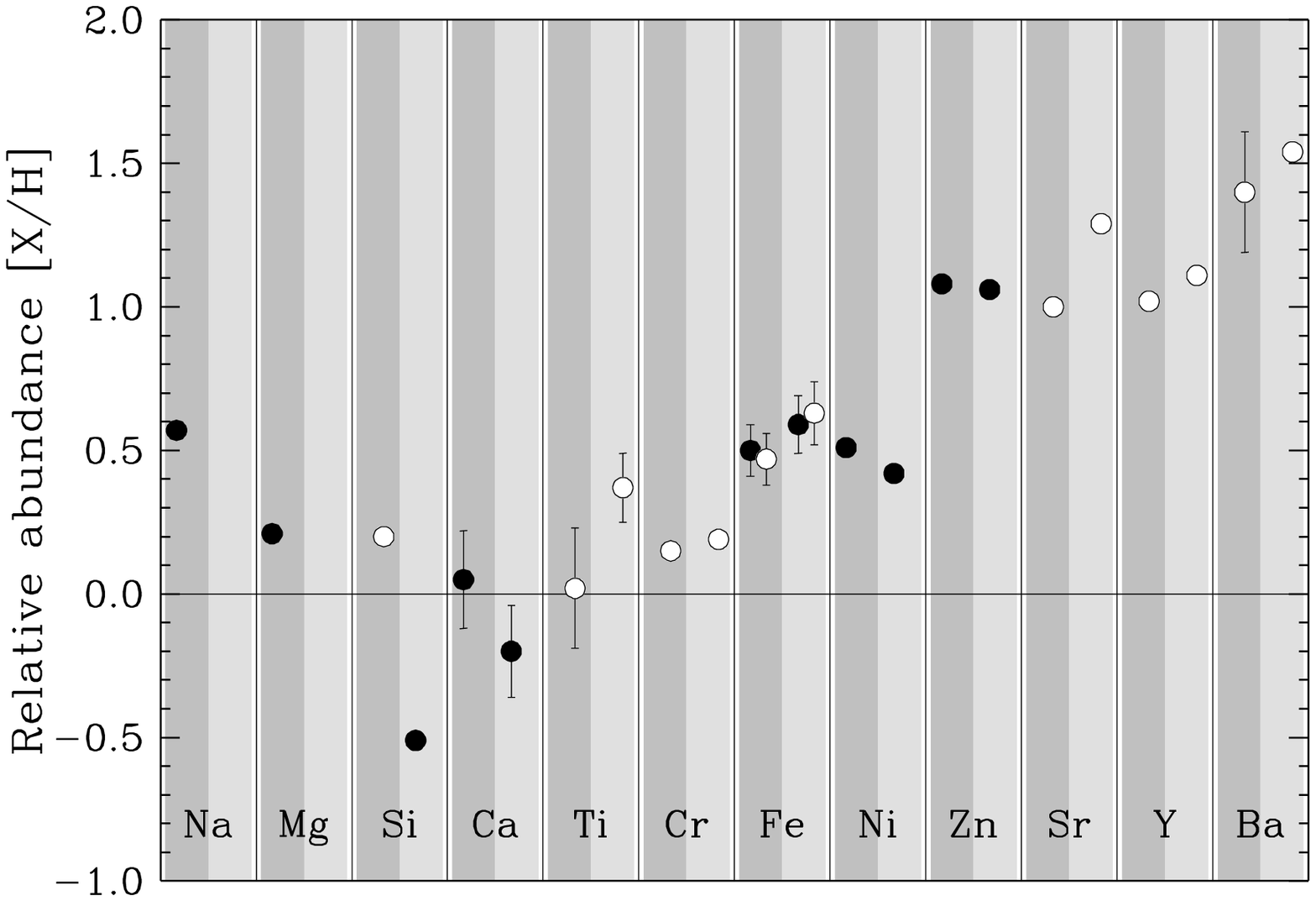}}
\epsfxsize=40mm
\epsfysize=40mm

\caption{Elemental abundances measured for \SW. Dark gray areas
correspond to the primary, and light gray to the secondary. Filled and
open symbols represent neutral and ionized species, respectively.}

\label{fig:swcma_abundances}
\end{figure}


\subsection{Photometric elements}
\label{sec:swcma_phel}

For this analysis we make use of the differential photometry for \SW\
in the Str\"omgren system that has been reported previously by Clausen
et al.\ (\cite{Clausen:08}). As described there, the $uvby$ light
curves consist of 820 observations in each band collected with the
0.5\,m Str\"omgren Automatic Telescope at ESO (La Silla), on 89 nights
during five periods between 1987 February and 1991 March.  The average
uncertainty per differential observation is about 5 mmag in $vby$ and
6 mmag in $u$, and the measurements scatter at these same levels
throughout all phases, indicating that the components of \SW\ are
constant within the precision of the data. Compared to the $UBV$ light
curves in the earlier study by Lacy (\cite{Lacy:97}), the present
observations are roughly twice as numerous and provide better orbital
coverage, particularly at the shoulders of the minima.

\begin{table}[t]
\caption{Adopted photometric elements for \SW.}
\label{tab:swcma_phel}
\begin{center}             
\begin{tabular}{lrrrr}             
\noalign{\smallskip}             
\hline\hline             
\noalign{\smallskip}             
$i$  (\degr)        & \multicolumn{4}{c}{ $88.59 \pm 0.20$} \\
$e\cos \omega$      & \multicolumn{4}{c}{$-0.30313 \pm 0.00009$}      \\
$e\sin \omega$      & \multicolumn{4}{c}{$+0.0883  \pm 0.0017$}       \\
$e$                 & \multicolumn{4}{c}{$ 0.3157  \pm 0.0004$}       \\
$\omega$  (\degr)   & \multicolumn{4}{c}{$163.76   \pm 0.29$}         \\
$r_p + r_s$         & \multicolumn{4}{c}{ $0.1718 \pm 0.0010$} \\       
$k$                 & \multicolumn{4}{c}{ $0.827\pm 0.018$} \\          
$r_p$               & \multicolumn{4}{c}{ $0.0940 \pm 0.0006$} \\       
$r_s$               & \multicolumn{4}{c}{ $0.0778 \pm 0.0013$} \\       
\noalign{\smallskip}             
\hline             
    
\noalign{\smallskip}             
                 & $y$~~    & $b$~~    & $v$~~   & $u$~~  \\           
\noalign{\smallskip}             
$J_s$            & 1.013  & 1.012  & 0.996  & 1.020\vspace{-0.8mm}  \\   
                 & $\pm3$ &$\pm2$  & $\pm3$ &$\pm 4$ \\   

$L_s/L_p$        & 0.693  & 0.692  & 0.681   & 0.697\vspace{-0.8mm} \\ 
                 &$\pm30$ &$\pm18$ &$\pm15$  &$\pm33$ \\   
\noalign{\smallskip}             
\hline             
\end{tabular}             
\end{center}            
\textsc{Note:}
The individual flux and luminosity ratios are based
on the mean stellar and orbital parameters.
\end{table}  

Light-curve solutions were performed with the JKTEBOP code (Southworth
et al.\ \cite{sms04a}, \cite{sms04b}), which is an updated and
expanded version of the original EBOP program based on the
Nelson-Davis-Etzel model (Nelson \& Davis \cite{nd72}; Etzel
\cite{e81}). This model represents the binary components as biaxial
ellipsoids and applies a simple bolometric reflection prescription,
but is perfectly adequate for the stars in \SW\ given their very small
deformation (oblateness of 0.0012 for the primary and 0.0008 for the
secondary, both much smaller than the tolerance for this model of
0.04; see Popper \& Etzel \cite{pe81}). Each of the $uvby$ light
curves was analyzed independently adopting the ephemeris given in
Sect.~\ref{sec:swcma}, with equal weight assigned to all observations.
The main parameters solved for are the radius ratio between the
secondary and the primary ($k \equiv r_s/r_p$), the sum of the radii
($r_p+r_s$), the inclination angle ($i$), the geometric factors
$e\sin\omega$ and $e\cos\omega$, the central surface brightness ratio
of the secondary in terms of the primary ($J_s$), a photometric scale
factor (the magnitude at quadrature), and a correction to the phase of
the primary minimum. The mass ratio was held fixed at the value
determined spectroscopically (Sect.~\ref{sec:swcma_spec_orb}).
Gravity darkening coefficients, $y_p$ and $y_s$, were computed from a
simple Planck approximation (see, e.g., Martynov \cite{m73}). Initial
solutions were carried out with limb-darkening (LD) coefficients
interpolated from theoretical calculations (van Hamme \cite{vh93}, and
ATLAS version of the Claret \cite{Claret:00} tables) according to the
effective temperatures and surface gravities for stars as determined
here.\footnote{Solar metallicity was adopted in interpolating all
limb-darkening coefficients for consistency in the comparison between
the different LD tables, given that the van Hamme (\cite{vh93})
calculations do not reach metallicities as high as those measured for
\SW. However, inspection of the Claret (\cite{Claret:00}) calculations
shows that the coefficients for [Fe/H] = 0.0 and [Fe/H] = +0.5 differ
very little.}

We found good agreement between the results from the $uvby$ passbands,
although small systematic differences were noticed between fits using
LD coefficients from different sources, as described in detail in
Appendix~\ref{ap:LDswcma} (available electronically). In view of this,
and out of concern that imposing LD coefficients from theory might
introduce subtle biases in the geometric elements, for the final fits
we chose to leave the LD coefficients free (linear law), with the only
constraint that they be equal for the primary and secondary given that
the temperatures are also very nearly the same. Additional experiments
with LD free were carried out to explore the effect of including third
light ($\ell_3$) as an extra parameter. For all four passbands we
found that $\ell_3$ was negligible compared to its error, and we
conclude third light is not significant.

\begin{figure}[t]
\resizebox{\hsize}{!}{\includegraphics{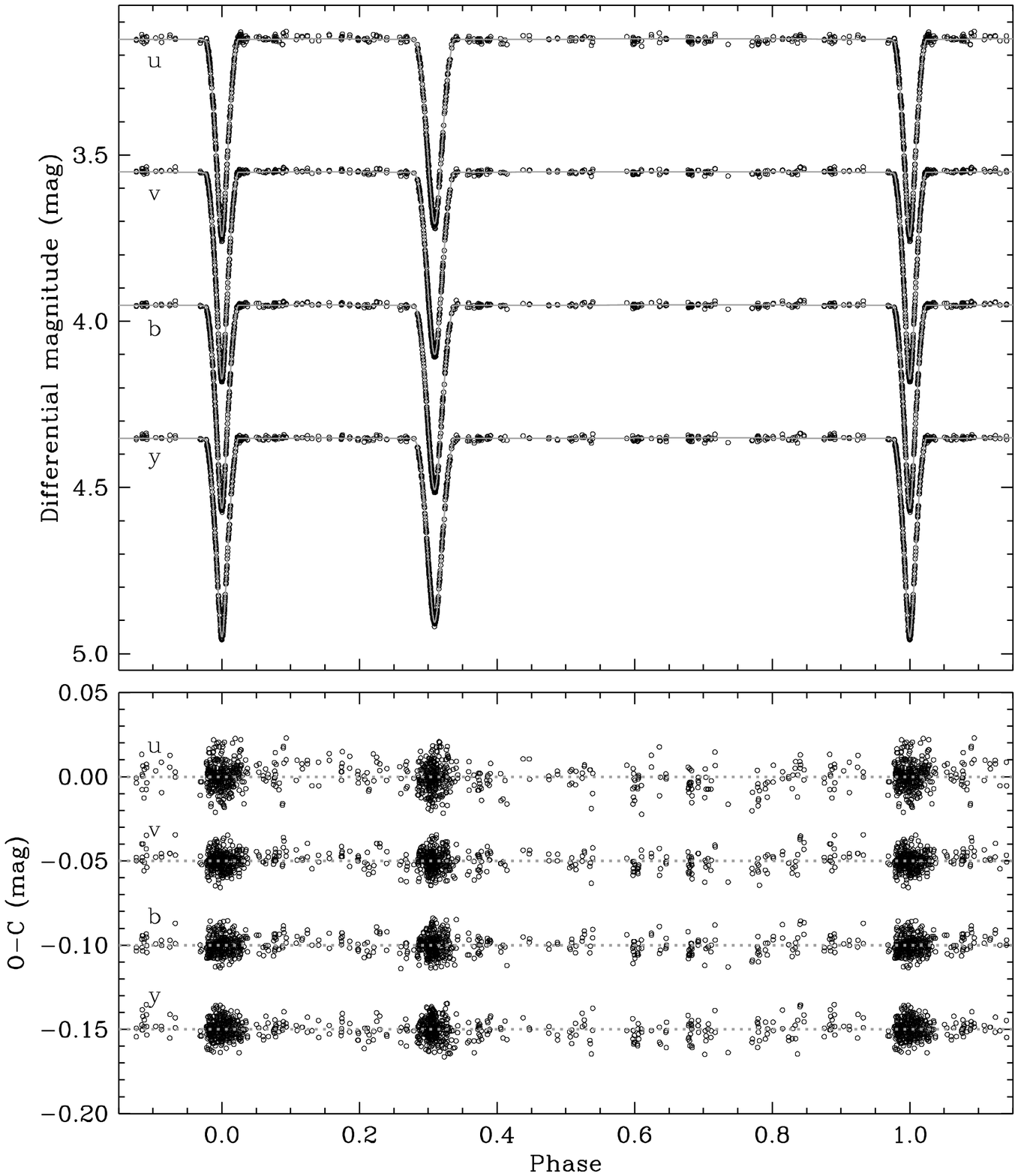}}
\caption{$uvby$ photometry for \SW\ together with our best fit models
(Table~\ref{tab:swcma_phel}). Phase 0.0 corresponds to the eclipse of
the more massive star (primary eclipse). Residuals are shown at the
bottom.}
\label{fig:swcma_figLC}
\end{figure}

The results are presented in Table~\ref{tab:swcma_phel}, in which the
values from the separate passbands have been averaged together, with
weights inversely proportional to the rms residual of each
solution. The uncertainties listed in the table were obtained by
running 1000 Monte Carlo simulations with JKTEBOP, and we consider
them to be more realistic than the formal (internal) errors from the
solutions presented in Appendix~\ref{ap:LDswcma}.  For the relative
radii we also computed uncertainties by propagating the errors in
$r_p+r_s$ and $k$ following Torres et al.\ (\cite{Torres:00}). This
procedure assumes the sum and ratio of the radii are uncorrelated, and
results in uncertainties that are more nearly equal (0.0011 and 0.0010
for $r_p$ and $r_s$, respectively) than those from Monte Carlo.
However, the Monte Carlo simulations indicate that in this case
$r_p+r_s$ and $k$ are in fact correlated, invalidating the error
propagation approach for \SW.  The radiative quantities in
Table~\ref{tab:swcma_phel} were computed from additional solutions in
which the geometry was held fixed to the weighted average from the
four bands. The eccentricity and longitude of periastron are more
precise than, but consistent with those obtained spectroscopically,
differing by 1.5$\sigma$ and 0.9$\sigma$, respectively. The light
ratios in $b$ and $y$ ($L_s/L_p = 0.69 \pm 0.02$ and $0.69 \pm 0.03$)
are in excellent agreement with the value from our spectroscopic
observations in Sect.~\ref{sec:swcma_spec_orb} ($0.68 \pm 0.04$ at
5187\,\AA).  According to the final light-curve elements, 72\% of the
light of the primary is blocked at the deeper minimum. The secondary
eclipse is nearly total: 97\% of the $y$-band light of that star lost
at that phase.

\begin{figure}
\resizebox{\hsize}{!}{\includegraphics{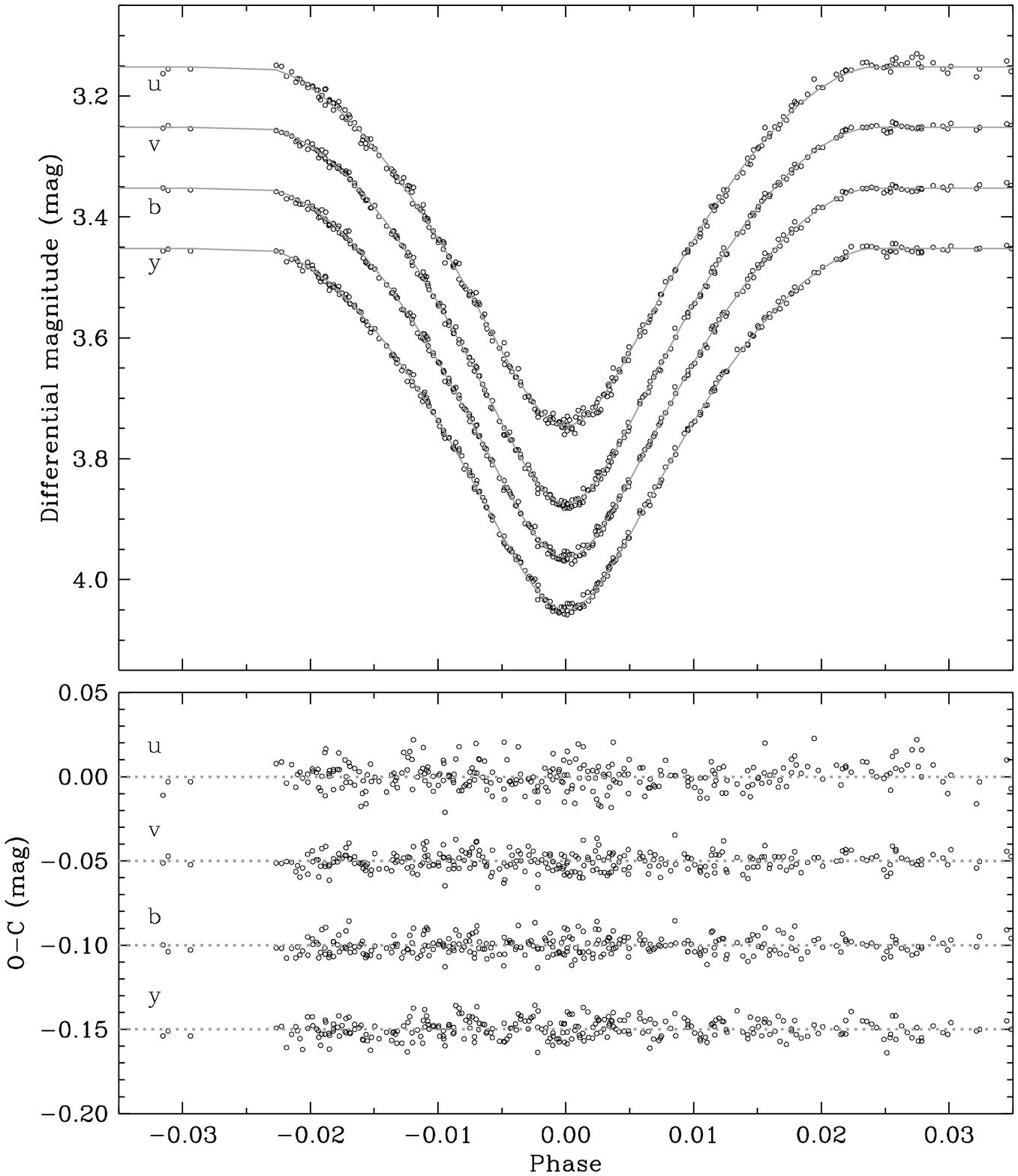}}
\caption{Enlarged view of Fig.~\ref{fig:swcma_figLC} around the primary
minimum of \SW.}
\label{fig:swcma_figLCprim}
\end{figure}


\begin{figure}[t]
\resizebox{\hsize}{!}{\includegraphics{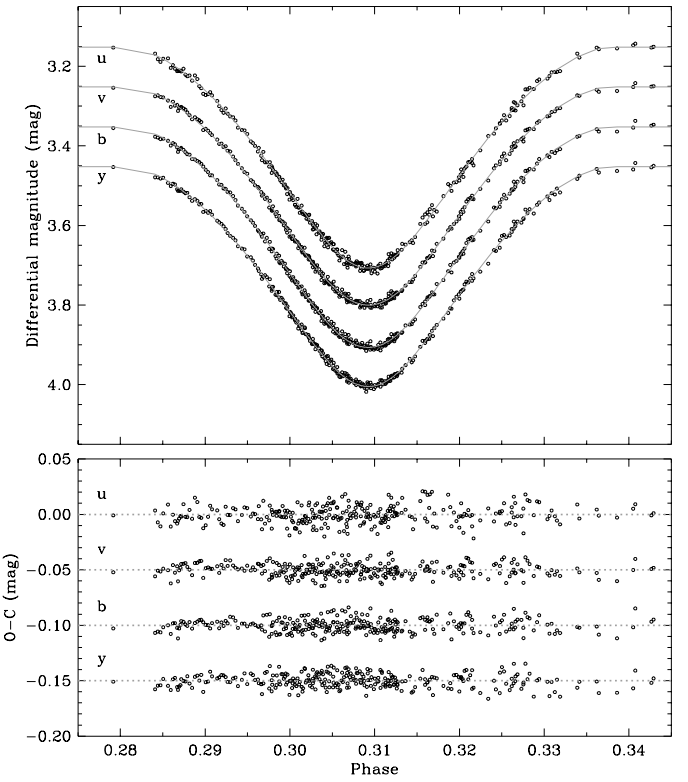}}
\caption{Enlarged view of Fig.~\ref{fig:swcma_figLC} around the secondary
minimum of \SW.}
\label{fig:swcma_figLCsec}
\end{figure}

The light curves along with our best fit models appear in
Fig.~\ref{fig:swcma_figLC}, and expanded views of the primary and
secondary minima are shown in Fig.~\ref{fig:swcma_figLCprim} and
Fig.~\ref{fig:swcma_figLCsec}. The $O-C$ residuals are displayed in the
bottom panels.


\subsection{Absolute dimensions}
\label{sec:swcma_absdim}

Our spectroscopy and differential photometry for \SW\ yield masses
with uncertainties of 0.6\% and 0.9\% for the primary and secondary,
and radii that are good to 0.7\% and 1.7\%, respectively. The values
of $M$ and $R$ are consistent with the analysis by Lacy
(\cite{Lacy:97}), but our errors are generally smaller (by about a
factor of three in the case of the masses).\footnote{We note that the
uncertainty of 0.5\,$M_{\sun}$ for the primary mass reported in Lacy's
Table~7 is likely a misprint; we have assumed it should read
0.05\,$M_{\sun}$.}  We list the masses and radii in
Table~\ref{tab:swcma_absdim}, along with other derived properties.

\begin{table}   
\caption{Astrophysical properties of the \SW\ system.}
\label{tab:swcma_absdim}
\begin{minipage}{\columnwidth}
\begin{center}    
\begin{tabular}{lrr} \hline\hline    
\noalign{\smallskip}    
                     &    Primary       &    Secondary      \\ 
\noalign{\smallskip}    
\hline    
\noalign{\smallskip}    
\multicolumn{3}{l}{Absolute dimensions:}                \\ 
~~~~$M$ ($M_{\sun}$)              &$2.239 \pm 0.014$  &$2.104 \pm 0.018$ \\ 
~~~~$R$ ($R_{\sun}$)              &$3.014 \pm 0.020$  &$2.495 \pm 0.042$  \\ 
~~~~$\log g$ (cgs)                & $3.8298 \pm 0.0065$ & $3.967 \pm 0.015$ \\
~~~~$v \sin i$ (\kms) $^{\mathrm{a}}$   & $24.0 \pm 1.5$      & $10.0 \pm 1.0$ \\ 
~~~~$v_{\rm sync}$ (\kms) $^{\mathrm{b}}$   & $15.1 \pm 0.1$    & $12.5 \pm 0.2$  \\ 
~~~~$v_{\rm psync}$ (\kms) $^{\mathrm{c}}$  & $24.5 \pm 0.2$    & $20.3 \pm 0.3$ \\ 
~~~~$v_{\rm peri}$ (\kms) $^{\mathrm{d}}$   & $30.6 \pm 0.2$    & $25.3 \pm 0.4$  \\ 
\noalign{\smallskip}
\multicolumn{3}{l}{Radiative and other properties:} \\
~~~~$T_{\mbox{\scriptsize eff}}$ (K)       &  $8200 \pm 150$    &   $8100 \pm 150$ \\
~~~~$M_{\mbox{\scriptsize bol}}$ (mag)   &  $0.817  \pm 0.081$  &   $1.281  \pm 0.088$ \\
~~~~$\log L/L_{\sun}$ & $ 1.566 \pm 0.032$ &    $ 1.380 \pm 0.035$ \\
~~~~$BC_V$  (mag)     & $ +0.02 \pm 0.10$  &    $ +0.02 \pm 0.10$ \\
~~~~$M_V$  (mag) &    $ 0.79 \pm 0.12$     &   $ 1.26 \pm 0.13$ \\
~~~~Distance  (pc)  & \multicolumn{2}{c}{$576 \pm 27$} \\
~~~~$V-M_V$  (mag)  & \multicolumn{2}{c}{$8.80 \pm 0.10$} \\
~~~~[Fe/H] $^{\mathrm{e}}$    & $+0.49 \pm 0.15$ & $+0.61 \pm 0.15$ \\
\noalign{\smallskip}
\multicolumn{3}{l}{Photometric indices:}    \\
~~~~$V$ $^{\mathrm{f}}$ &         $ 9.721 \pm 0.021$  &        $10.119 \pm 0.029$\\              
~~~~$(b-y)$ $^{\mathrm{f}}$ &     $ 0.093 \pm 0.012$  &        $ 0.095 \pm 0.014$\\
~~~~$m_1$ $^{\mathrm{f}}$   &     $ 0.203 \pm 0.012$  &        $ 0.220 \pm 0.013$\\
~~~~$c_1$ $^{\mathrm{f}}$   &     $ 1.031 \pm 0.015$  &        $ 0.987 \pm 0.020$\\
~~~~$E(b-y)$    & \multicolumn{2}{c}   {$0.023 \pm 0.010$} \\
\noalign{\smallskip}            
\hline
\noalign{\smallskip}
\end{tabular}            
\begin{list}{}{}
\item[$^{\mathrm{a}}$] Measured projected rotational velocity.
\item[$^{\mathrm{b}}$] $v \sin i$ expected for synchronous rotation.
\item[$^{\mathrm{c}}$] $v \sin i$ expected for pseudo-synchronous rotation.
\item[$^{\mathrm{d}}$] Expected $v \sin i$ if synchronized at periastron.
\item[$^{\mathrm{e}}$] Values representative of the surface layers only.
\item[$^{\mathrm{f}}$] Not corrected for interstellar absorption/reddening.
\end{list}
\end{center}
\textsc{Note:} The bolometric corrections ($BC_V$) are adopted from
Flower (\cite{flower96}), along with with $T_{\rm eff\sun} = 5780$ K,
and $M_{bol\sun} = 4.73$ (see Torres \cite{Torres:10b}). An additional
error contribution of 0.10~mag is added in quadrature to the $BC_V$
uncertainty propagated from the temperature uncertainties.
\end{minipage}
\end{table}                                  

Interstellar reddening may be estimated from the $uvby\beta$ indices
out of eclipse reported by Clausen et al.\ (\cite{Clausen:08}), which
agree well with those of Wolf \& Kern (\cite{Wolf:83}). Using the
calibration by Crawford (\cite{Crawford:79}) we obtain $E(b-y) =
0.023$, to which we assign a conservative uncertainty of 0.01 mag. We
adopt this value in the following. The corresponding visual extinction
is $A(V) = 0.10 \pm 0.04$.  Other estimates of the extinction in the
direction of \SW\ may be inferred from dust maps (which measure the
\emph{total} extinction), after proper correction for the distance to
the system by iterations. However, we find those estimates to be
larger and rather discrepant: from Hakkila et al.\ (\cite{Hakkila:97})
we obtain $A(V) = 0.16$, corresponding to $E(b-y) = 0.037$, while the
maps of Schlegel et al.\ (\cite{Schlegel:98}) suggest $A(V) = 0.32$,
or $E(b-y) = 0.077$, at the distance to the system.

The combined $uvby$ indices from Clausen et al.\ (\cite{Clausen:08})
and the light ratios from our photometric solutions yield the
individual indices for the components, also listed in
Table~\ref{tab:swcma_absdim}. The uncertainties include the
contribution from all observational errors.  The individual bolometric
luminosities follow from the spectroscopic temperatures and the radii,
and the distance was obtained from the luminosities, the de-reddened
$V$ magnitude, and bolometric corrections from Flower
(\cite{flower96}). We infer a distance to the system of $576 \pm
27$\,pc, and separate calculations for the primary and secondary agree
closely (giving $583 \pm 36$\,pc and $566 \pm 37$\,pc, respectively),
which is an indication of good internal consistency.  Knowledge of the
individual photometric indices for the components provides an
opportunity to derive temperature estimates as a check on the
spectroscopic values, to the extent that existing temperature
calibrations for normal A-type stars can be applied to objects
displaying the Am phenomenon.  This appears to be the case, according
to Netopil et al.\ (\cite{Netopil:08}).  After correcting the
individual indices for the effects of interstellar reddening,
application of the calibration of Napiwotzki et al.\ (\cite{nsw93})
leads to estimates of $8188 \pm 157$\,K and $8168 \pm 173$\,K for the
primary and secondary, in good accord with the spectroscopic values. A
0.02~mag error in $b-y$ changes these results by about 200\,K.  Using
the central surface brightness ratio $J_s$ in the $y$ band obtained
from our light curve solutions with the flux calibration by Popper
(\cite{dmp80}) yields an estimate of the temperature difference
between the stars of $\Delta T_{\rm eff} = -32$\,K, suggesting the
secondary (less massive star) is marginally hotter than the primary,
though well within the uncertainties of our other determinations.

The $v \sin i$ values discussed in Sect.~\ref{sec:swcma_spec_orb} that
provide the best fit to the CfA spectra are $29 \pm 3$~\kms\ and $15
\pm 2$~\kms\ for the primary and secondary, respectively. Strictly
speaking, however, these measures represent the {\it total\/} line
broadening, and not just rotation, and indeed they are systematically
larger than those obtained from the FIES spectrum ($24.0 \pm
1.5$~\kms\ and $10.0 \pm 1.0$~\kms). This is most likely due to the
fact that the synthetic templates used to analyze the CfA spectra have
been calculated with a radial-tangential macroturbulent velocity
appropriate for solar-type stars ($\zeta_{\rm RT} = 1.5$~\kms),
whereas the values of $\zeta_{\rm RT}$ expected for A-type stars (at
least normal ones) are significantly higher (see, e.g., Gray
\cite{Gray:05}). Indeed, for the FIES analysis we adopted $\zeta_{\rm
RT} = 8$~\kms. We therefore rely on the FIES determination of $v \sin
i$, which we report in Table~\ref{tab:swcma_absdim}. Also listed there
are the $v \sin i$ values predicted if the stars had their spins
synchronized with the mean orbital motion ($v_{\rm sync}$), with the
motion at periastron ($v_{\rm peri}$), and with an intermediate
equilibrium rate of `pseudo-synchronization' (see Hut
\cite{Hut:81}). The comparison would suggest the primary component of
\SW\ is rotating at the pseudo-synchronous rate and the secondary is
spinning considerably more slowly than pseudo-synchronous.
We note here also that Lacy (\cite{Lacy:97}) measured rather different
values for $v \sin i$ of $30 \pm 2$~\kms\ for the primary and $21 \pm
3$~\kms\ for the secondary, which would lead to different conclusions
regarding synchronization.

The Am nature of both stars is supported by the abundance
pattern we see, most notably that the heavy elements are overabundant
by an order of magnitude or more compared to solar. Enhancements 
seem rather similar for the two components, with a few exceptions
noted earlier.


\section{HW\,CMa}
\label{sec:hwcma}

The long-period eclipsing binary \HW\ (HD\,54549, $V = 9.19$, A6, $P =
21.1$\,days) was discovered serendipitously by Liu et al.\
(\cite{Liu:92}) in the course of spectroscopic monitoring of \SW,
which is only about 2\farcm5 away. Beyond the initial estimates of the
minimum masses by these authors (1.74\,$M_{\sun}$ and
1.80\,$M_{\sun}$), no determinations of the absolute dimensions have
been reported in the literature. Accurate differential $uvby$
photometry of \HW\ was published by Clausen et al.\
(\cite{Clausen:08}), who confirmed that the combination of high
orbital eccentricity ($e \approx 0.50$), low inclination ($i \approx
84\fdg7$), and a line of apsides practically aligned with the line of
sight result in the complete absence of eclipses near apastron. The
shallow (0.13~mag) eclipses that do take place near periastron
correspond to the less massive star being occulted by the other.
In the following we refer to these as the primary eclipses,
following the convention adopted by Clausen et al.\
(\cite{Clausen:08}). Secondary eclipses would be expected at phase
$\sim$0.52, but as mentioned above, they do not occur.

An accurate ephemeris has been determined by Clausen et al.\
(\cite{Clausen:08}) as
\begin{displaymath}
\label{eq:hwcma_eph}
{\rm Min~I~(HJD)} = 2\,452\,279.6787(4) + 21\fd1178329(33) \times E~,
\end{displaymath}
which we adopt here. Apsidal motion is expected in view of the large
eccentricity, but due to the lack of secondary eclipses this can
only be measured spectroscopically.


\subsection{Radial velocities and spectroscopic orbit}
\label{sec:hwcma_spec_orb}

\begin{figure}
\resizebox{\hsize}{!}{\includegraphics{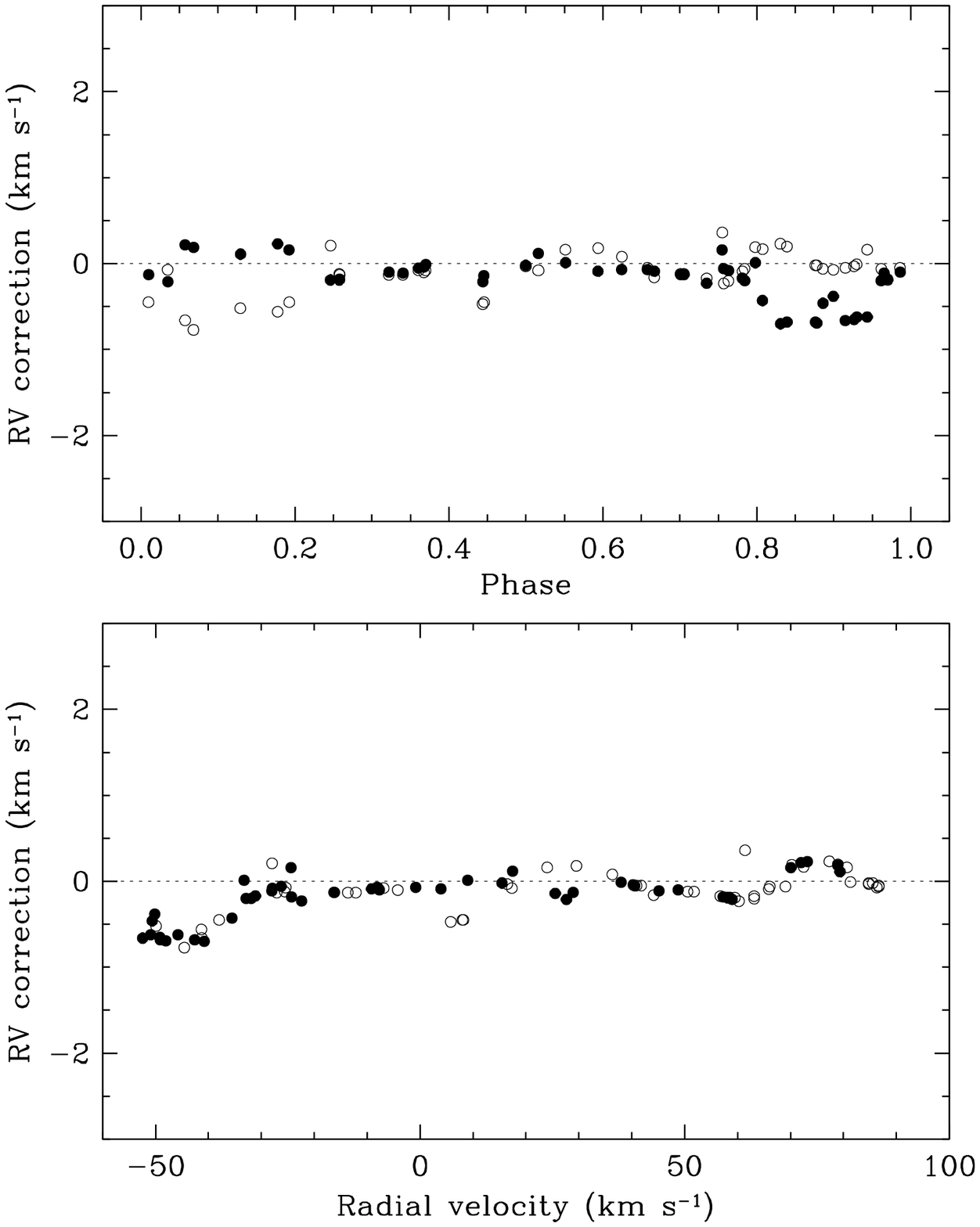}}
\epsfxsize=80mm
\caption{Systematic errors in the raw TODCOR velocities of \HW\
determined from simulations based on synthetic binary spectra. Filled
circles correspond to the more massive star, and open circles to the
other component.  The velocity differences are shown as a function of
orbital phase and radial velocity, and have been applied to the raw
velocities as corrections. Phase 0.0 corresponds to the eclipse of the
less massive star.}
\label{fig:hwcorr}
\end{figure}

Spectroscopic observations of \HW\ were gathered between 1988 November
and 1999 March with the same instrumentation used for \SW\ and
described in Sect.~\ref{sec:swcma_spec_orb}. A total of 48 spectra
were included in the analysis, with S/N ratios ranging from 9 to 39
per resolution element of 8.5~\kms. This is a superset of the
observations reported by Liu et al.\ (\cite{Liu:92}), who used the
same spectrographs.  The optimization of the synthetic templates and
determination of radial velocities using TODCOR follows the procedures
described earlier. The surface gravities for the templates were held
fixed at $\log g = 4.0$ for both stars, which is the value in our grid
of synthetic spectra closest to the final estimates reported in
Sect.~\ref{sec:hwcma_absdim}. As was the case for \SW,
cross-correlation grids with TODCOR indicated a preference for a
metallicity of [Fe/H] = +0.50 for both stars, the highest available in
our library of templates, but in this case we saw no obvious
difference between the two stars. A value this high is again a hint of
the possible Am nature of the components.  The template parameters
adopted for the RV determinations are $T_{\rm eff} = 7500$\,K (less
massive star, the `primary') and $T_{\rm eff} = 7750$\,K (more massive
`secondary'), and $v \sin i = 16$~\kms\ for both components.

\begin{figure}
\resizebox{\hsize}{!}{\includegraphics{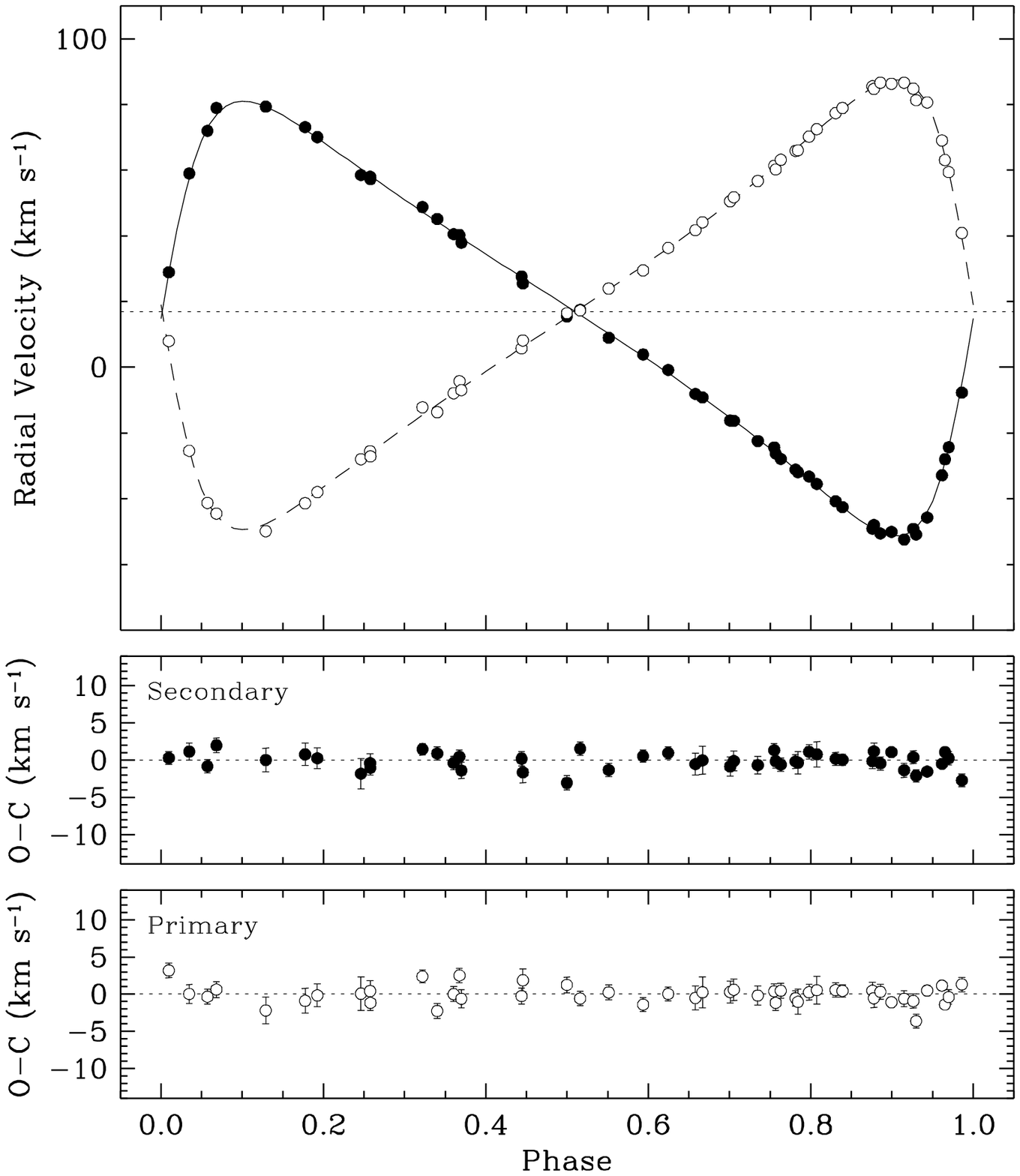}}
\caption{Measured radial velocities for \HW\ and orbital solution
(solid line for the more massive `secondary' star, dashed for the
primary). The dotted line in the top panel indicates the
center-of-mass velocity of the binary, and phase 0.0 corresponds to
eclipse of the less massive component. Residuals are shown at the
bottom.}
\label{fig:hworbit}
\end{figure}

Simulations with synthetic binary spectra were run in the same way as
for \SW\ to estimate and correct for systematic errors in the raw
velocities resulting from lines shifting in and out of our narrow
spectral window with orbital phase. The magnitude of these effects for
\HW\ is less than 1~\kms\ (see Fig.~\ref{fig:hwcorr}). Nevertheless,
the impact on the masses and radii is not negligible in this case: the
masses increase by about 1\% ($\sim$1.5 times their final errors), and
the radii by slightly more than 0.3\%. The final velocities including
these corrections are listed in Table~\ref{tab:hwcma_rv}, and
incorporate also the adjustments for instrumental shifts from run to
run described in Sect.~\ref{sec:swcma_spec_orb}. The RV measurements
and our best-fit spectroscopic orbital solution are shown graphically
in Fig.~\ref{fig:hworbit}, and the elements are given in
Table~\ref{tab:hwcma_orb}. The light ratio inferred from the CfA
spectra using TODCOR is $L_p/L_s = 0.92 \pm 0.04$ (primary/secondary),
which corresponds to the mean wavelength of those observations.

\begin{table}
\caption{Spectroscopic orbital solution for \HW.}
\label{tab:hwcma_orb} 
\begin{center}
\begin{tabular}{lr}
\hline\hline\noalign{\smallskip}
\multicolumn{1}{c}{Parameter} &
\multicolumn{1}{c}{Value} \\
\noalign{\smallskip}\hline\noalign{\smallskip}
\multicolumn{2}{l}{Adjusted quantities:}  \\
~~~~$K_p$ (\kms)             & $ 68.47 \pm 0.23 $  \\
~~~~$K_s$ (\kms)             & $ 66.18 \pm 0.21 $  \\
~~~~$\gamma$ (\kms)          & $+16.91 \pm 0.10 $  \\
~~~~$e$                      & $ 0.5016 \pm 0.0018 $ \\
~~~~$\omega$ $(\degr)$ $^{\mathrm{a}}$      & $ 86.38 \pm 0.26 $ \\
\noalign{\smallskip}
\multicolumn{2}{l}{Adopted quantities:}  \\
~~~~$P$ (days)                      & 21.1178329\\
~~~~$T_{\rm I}$ (HJD$-$2\,400\,000) $^{\mathrm{b}}$        &  52279.6787 \\
\noalign{\smallskip}
\multicolumn{2}{l}{Derived quantities:}  \\
~~~~$M_p \sin^3i$ ($M_{\sun}$)      & $ 1.700  \pm 0.011  $ \\
~~~~$M_s \sin^3i$ ($M_{\sun}$)      & $ 1.759  \pm 0.012  $ \\
~~~~$q \equiv M_p/M_s $             & $ 0.9665 \pm 0.0043 $ \\
~~~~$a_p \sin i$ ($10^6$~km)        & $17.202  \pm 0.055  $ \\
~~~~$a_s \sin i$ ($10^6$~km)        & $16.626  \pm 0.050  $ \\
~~~~$a \sin i$ ($R_{\sun}$)         & $48.63   \pm 0.10   $ \\
\noalign{\smallskip}
\multicolumn{2}{l}{Other quantities pertaining to the fit:}  \\
~~~~$N_{obs}$                      &    48 \\
~~~~Time span (days)               &  3759 \\
~~~~$\sigma_p$ (\kms)              &  1.05 \\
~~~~$\sigma_s$ (\kms)              &  0.95 \\
\noalign{\smallskip}
\hline
\end{tabular}
\begin{list}{}{}
\item[$^{\mathrm{a}}$] Longitude of periastron for the less massive (primary) star.
\item[$^{\mathrm{b}}$] Time of central primary eclipse (eclipse of the less massive star).
\end{list}
\end{center}
\end{table}

The final surface gravities of the stars ($\log g \approx 4.24$;
Sect.~\ref{sec:hwcma_absdim}) are approximately halfway between two
steps in our library of templates, and our velocity determinations
above used the lower of the two values ($\log g = 4.0$). Given that
effective temperatures are strongly correlated with surface gravity in
determinations based on our CfA spectra, a more accurate temperature
may be derived by interpolation to $\log g = 4.24$. In this way we
find $T_{\rm eff} = 7630 \pm 150$\,K and $7780 \pm 150$\,K for the
primary and secondary, and $v \sin i$ values (including other sources
of line broadening; see Sect.~\ref{sec:hwcma_absdim}) of 15~\kms\ for
both stars. We expect the temperatures to be more accurate than in the
case of \SW, as the line strengths of the two stars in \HW\ appear
more or less equally enhanced compared to solar metallicity stars.


\subsection{Chemical abundances}
\label{sec:hwcma_abund}

For a detailed study of the chemical composition of \HW\ a spectrum of
the binary was obtained on UT 2008 February 26 with the FIES
instrument on the 2.5\,m Nordic Optical Telescope (La Palma). The
Julian date at mid exposure is HJD 2\,454\,523.38203, corresponding to
orbital phase 0.2469. The S/N ratio of this observation is
approximately 120 per pixel at a wavelength of 6160\,\AA. The
reduction and analysis were carried out in the same way as described
in Sect.~\ref{sec:swcma_abund}. For the VWA analysis we selected only
lines with measured equivalent widths smaller than 65\,m\AA\ in the
composite spectrum (corresponding to about 140\,m\AA\ for the primary
and 120\,m\AA\ for the secondary), and we assumed a light ratio of
$L_p/L_s = 0.90$ for the $V$ band (see Sect.~\ref{sec:hwcma_absdim}).
The surface gravities of the components were both held fixed at a
preliminary estimate of $\log g = 4.23$, near the final values in
Sect.~\ref{sec:hwcma_absdim}. The $v \sin i$ values were measured to
be $12.0 \pm 1.0$\,\kms\ for both stars (for an adopted
macroturbulent velocity of $\zeta_{\rm RT} = 7$\,\kms), although this
has little impact on the abundance results. The temperatures yielding
the best agreement between the abundances of \ion{Fe}{i} and
\ion{Fe}{ii} are 7500\,K for the primary and 7700\,K for the
secondary, with uncertainties estimated at 150\,K.

\begin{table}
\caption{Abundances for the components of \HW.}
\label{tab:hwcma_abundances}
\begin{center}
\begin{tabular}{lcccccc}
\hline\hline\noalign{\smallskip}
& \multicolumn{3}{c}{Primary} & \multicolumn{3}{c}{Secondary} \\
\noalign{\smallskip}\hline\noalign{\smallskip}
Element & [X/H] & $\sigma$ & $N$ & [X/H] & $\sigma$ & $N$ \\
\noalign{\smallskip}\hline\noalign{\smallskip}
\ion{C}{i}   &  $-$0.66 &         &   1 &          &        &      \\
\ion{Na}{i}  &    +0.34 &         &   1 &    +0.37 &        &   1  \\
\ion{Mg}{i}  &          &         &     &  $-$0.24 &        &   2  \\
\ion{Si}{i}  &    +0.02 & 0.09    &   6 &    +0.13 & 0.12   &   4  \\
\ion{Si}{ii} &    +0.40 &         &   1 &    +0.24 &        &   2  \\
\ion{Ca}{i}  &  $-$0.64 & 0.17    &   4 &  $-$0.65 &        &   2  \\
\ion{Ti}{ii} &    +0.25 &         &   2 &    +0.18 & 0.15   &   4  \\
\ion{Cr}{i}  &    +0.08 &         &   1 &  $-$0.31 &        &   1  \\
\ion{Cr}{ii} &    +0.57 & 0.18    &   3 &    +0.35 & 0.11   &   6  \\
\ion{Fe}{i}  &    +0.30 & 0.08    &  31 &    +0.29 & 0.08   &  24  \\
\ion{Fe}{ii} &    +0.37 & 0.09    &  12 &    +0.35 & 0.10   &  15  \\
\ion{Ni}{i}  &    +0.32 & 0.15    &   5 &    +0.52 & 0.09   &  12  \\
\ion{Zn}{i}  &    +0.28 &         &   1 &    +0.26 &        &   2  \\
\ion{Sr}{ii} &    +0.72 &         &   1 &    +0.10 &        &   1  \\
\ion{Y}{ii}  &    +0.87 &         &   1 &    +0.75 &        &   1  \\
\ion{Ba}{ii} &    +1.14 &         &   2 &    +1.16 & 0.18   &   3  \\
\noalign{\smallskip}\hline
\end{tabular}
\end{center}
\textsc{Note:} These abundances are based on $v_{\rm mic} =
2.55$~\kms\ for both stars, and [Fe/H] = +0.30 for the model
atmospheres applied in the analysis. \ion{Fe}{i} abundances include
corrections for NLTE. $N$ represents the number of lines of each
element, and $\sigma$ is the scatter of the measurements (listed only
if $N > 2$). The primary is the less massive star in the binary.
\end{table}

We tested a range of [Fe/H] values for the model atmospheres used in
the VWA analysis, as well as different values of the microturbulence
in order to examine the sensitivity of the results. The final [Fe/H]
level of +0.30 dex for the model atmospheres was chosen to be
consistent with that found below from the equivalent widths.
Microturbulent velocities were $2.55 \pm 0.80$\,\kms\ for both
stars. Individual elemental abundances for 16 species are presented in
Table~\ref{tab:hwcma_abundances} on the scale of the solar abundances
of Grevesse et al.\ (\cite{Grevesse:07}). An NLTE correction to
\ion{Fe}{i} of +0.06~dex for the primary and +0.07~dex for the
secondary (Rentzsch-Holm \cite{Rentzsch-Holm:96}) is included.  The
average iron abundance is essentially identical for the two stars:
[Fe/H] $= +0.33 \pm 0.15$ for the primary and $+0.32 \pm 0.15$ for the
secondary (uncertainties include contributions from errors in
$T_{\rm eff}$ and $v_{\rm mic}$).

\begin{figure}
\resizebox{\hsize}{!}{\includegraphics[angle=0]{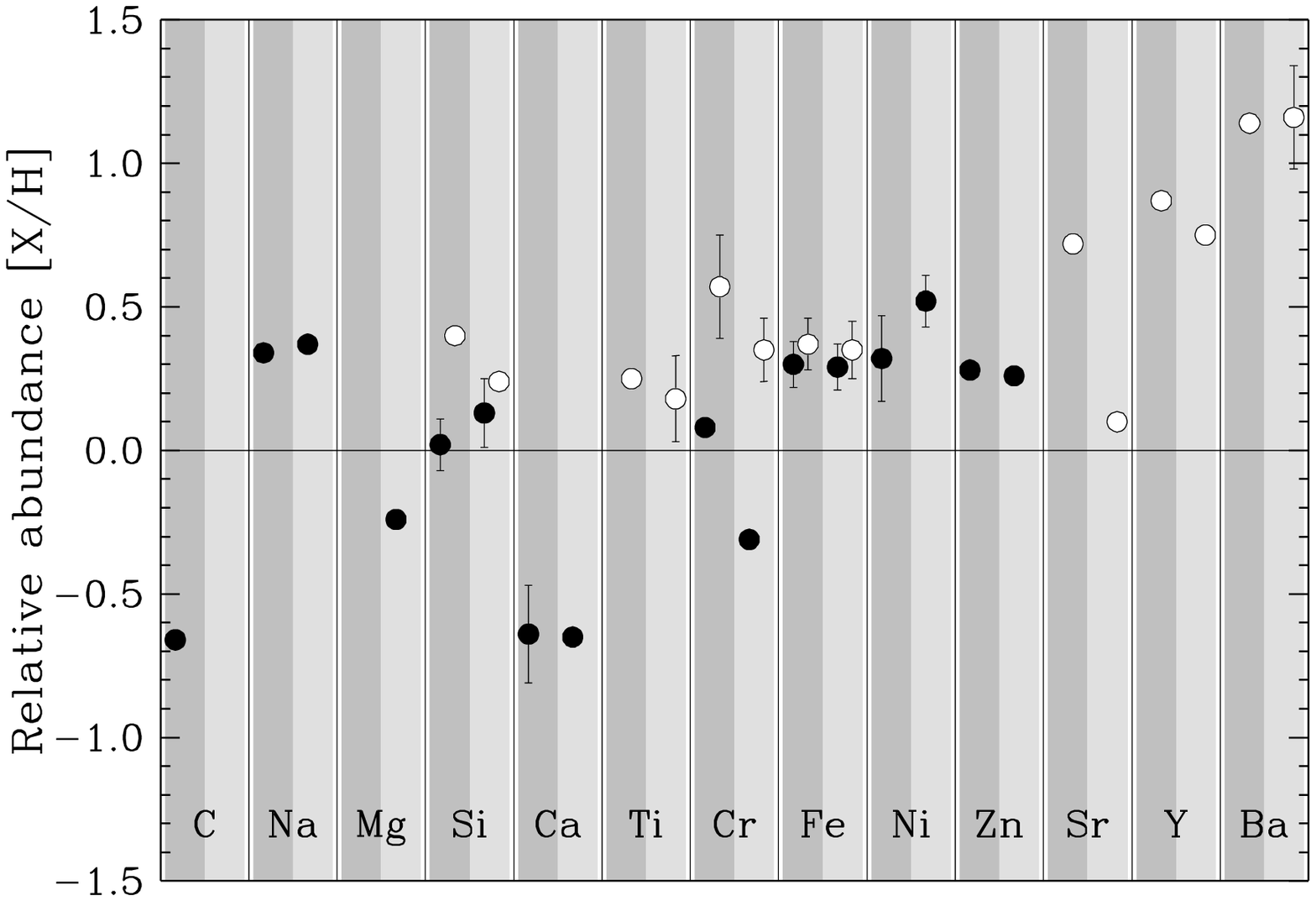}}

\caption{Elemental abundances measured for \HW. Dark gray areas
correspond to the primary (less massive star), and light gray to
the secondary. Filled and open symbols represent neutral and ionized
species, respectively.}

\label{fig:hwcma_abundances}
\end{figure}

The abundance pattern for the components of \HW\ is shown graphically
in Fig.~\ref{fig:hwcma_abundances}. There are strong indications
of the Am phenomenon in both stars: Ca seems characteristically
deficient, along with C (at least in the primary), and heavier
elements appear to be enhanced. In particular, Sr, Y, and Ba are
generally much stronger than normal.  The Fe/Ca difference is the same
for both components (0.97 dex), and is considerably larger than we
found for \SW. Additionally, for that system we measured Zn to be
stronger than iron by a factor of 3--4 in both stars, while in \HW\
these two elements have about the same abundance. Scandium is a
highly diagnostic element in Am stars, where it is usually found to be
deficient. Unfortunately, however, no lines of this element are
suitable for measurement in our spectra of either system.
Nevertheless, the overall abundance pattern in both \SW\ and \HW\ is
quite similar to that found in many other studies of Am stars (see,
e.g., Okyudo \& Sadakane \cite{Okyudo:90}, Hui-Bon-Hoa \cite{Hui:00},
Yushchenko et al.\ \cite{Yushchenko:04}, Iliev et al.\
\cite{Iliev:06}, Fossati et al.\ \cite{Fossati:07}, Gebran et al.\
\cite{Gebran:08}, \cite{Gebran:10}).


\subsection{Photometric elements}
\label{sec:hwcma_phel}

The $uvby$ differential photometry used for \HW\ is that reported by
Clausen et al.\ (\cite{Clausen:08}), obtained with the 0.5\,m
Str\"omgren Automatic Telescope at ESO (La Silla) between 1989
February and 2002 March. A total of 415 observations were obtained in
each filter, with similar average uncertainties per differential
measurement as for \SW\ ($\sim$5 mmag in $vby$ and $\sim$6 mmag in
$u$). The system was observed intensively at the expected phase of
secondary eclipse, but no significant drop in brightness was detected,
as mentioned earlier. With only a single eclipse to constrain the
geometry, this data set therefore poses special challenges to extract
reliable orbital elements: the central surface brightness ratio $J_s$
is completely unconstrained, there is no information in the light
curve on the eccentricity ($e \cos\omega$), and the radius ratio $k$
is very poorly defined without an external constraint. 

As for the case of \SW, we used the JKTEBOP program for the
light-curve solutions given that the stars are essentially spherical
(oblateness $\sim 0.00006$). The ephemeris adopted is that of
Sect.~\ref{sec:hwcma}. The eccentricity and longitude of periastron
were held fixed at the spectroscopic values, as was the mass ratio.
In order to constrain $J_s$ we calculated model spectra for each star
from the MARCS series (Gustafsson et al.\ \cite{Gustafsson:08}) for
the adopted temperatures (7560\,K and 7700\,K for the primary and
secondary; see Sect.~\ref{sec:hwcma_absdim}), and used them to obtain
$J_s$ estimates of 1.078 ($u$), 1.105 ($v$), 1.091 ($b$), 1.075 ($y$),
and 1.081 (5187\,\AA), with only a small dependence on [Fe/H] and
$\log g$. The $y$ value is nearly identical to that obtained from the
empirical flux scale of Popper (\cite{dmp80}), $J_s = 1.076$ in $V$,
supporting the accuracy of the theoretical calculations.

Because of its sensitivity to $k$, the spectroscopic light ratio
($L_s/L_p \propto J_s k^2$) provides a very useful constraint on the
poorly defined ratio of the radii in \HW. The average spectroscopic
light ratio from three independent determinations described below in
Sect.~\ref{sec:hwcma_absdim} is $L_p/L_s = 0.90 \pm 0.02$, or $L_s/L_p
= 1.11 \pm 0.02$. This corresponds strictly to the mean wavelength of
our spectroscopic observations (5187\,\AA).  Light ratios in the
Str\"omgren bands that are needed for the light curve fits were
inferred by making an initial estimate of $k$ requiring that the
combination of the measured $L_s/L_p$ and $k$ match the theoretical
value of $J_s$ at 5187\,\AA. We then used this estimate of $k = 1.013$
along with the $J_s$ values in $uvby$ to obtain the corresponding
light ratios: 1.106 ($u$), 1.134 ($v$), 1.119 ($b$), and 1.103 ($y$).

\begin{table}            
\caption{Adopted photometric elements for \HW.
}
\label{tab:hwcma_phel}
\begin{center}             
\begin{tabular}{lrrrr}             
\noalign{\smallskip}             
\hline\hline             
\noalign{\smallskip}             
$i$  (\degr)        & \multicolumn{4}{c}{ $84.84 \pm 0.06$} \\
$r_p + r_s$         & \multicolumn{4}{c}{ $0.06769 \pm 0.00066$} \\       
$k$                 & \multicolumn{4}{c}{ $1.012\pm 0.013$} \\          
$r_p$               & \multicolumn{4}{c}{ $0.03365 \pm 0.00037$} \\       
$r_s$               & \multicolumn{4}{c}{ $0.03404 \pm 0.00042$} \\       
\noalign{\smallskip}             
\hline             

\noalign{\smallskip}             
                 & $y$~~    & $b$~~    & $v$~~   & $u$~~  \\           
\noalign{\smallskip}             
$J_s$            & 1.075  & 1.091  & 1.105  & 1.078\vspace{-0.8mm}  \\   
                 & $\pm2$ &$\pm2$  & $\pm2$ &$\pm 2$ \\   

$L_s/L_p$        & 1.100  & 1.121  & 1.135   & 1.108\vspace{-0.8mm} \\ 
                 &$\pm20$ &$\pm20$ &$\pm21$  &$\pm20$ \\   
\noalign{\smallskip}             
\hline             
\end{tabular}             
\end{center}            
\textsc{Note:} Flux ratios are based on the measured effective
temperatures and MARCS models.  Luminosity ratios are constrained to
match the spectroscopic value at 5187\,\AA. The geometric elements are
the weighted mean of the $vby$ values from
Table~\ref{tab:hwcma_ebop_vh} using linear limb-darkening coefficients
from van Hamme (\cite{vh93}).

\end{table}  

JKTEBOP fits were performed separately in each passband, fixing $J_s$
and $L_s/L_p$ to the values specified above and holding $e \sin\omega$
and $e \cos\omega$ fixed as well (from the spectroscopy). We solved
for $i$, $r_p+r_s$, $k$, and the usual photometric scale factor and
phase offset.  Solutions were carried out using LD coefficients for
the linear law from both the Claret (\cite{Claret:00}) and van Hamme
(\cite{vh93}) tables. The geometric elements with each set of
coefficients show good agreement between the $v$, $b$, and $y$ bands,
with $u$ being more discrepant (and also more uncertain). We describe
the results of these tests in Appendix~\ref{ap:LDhwcma} (available
electronically).

\begin{figure}
\resizebox{\hsize}{!}{\includegraphics{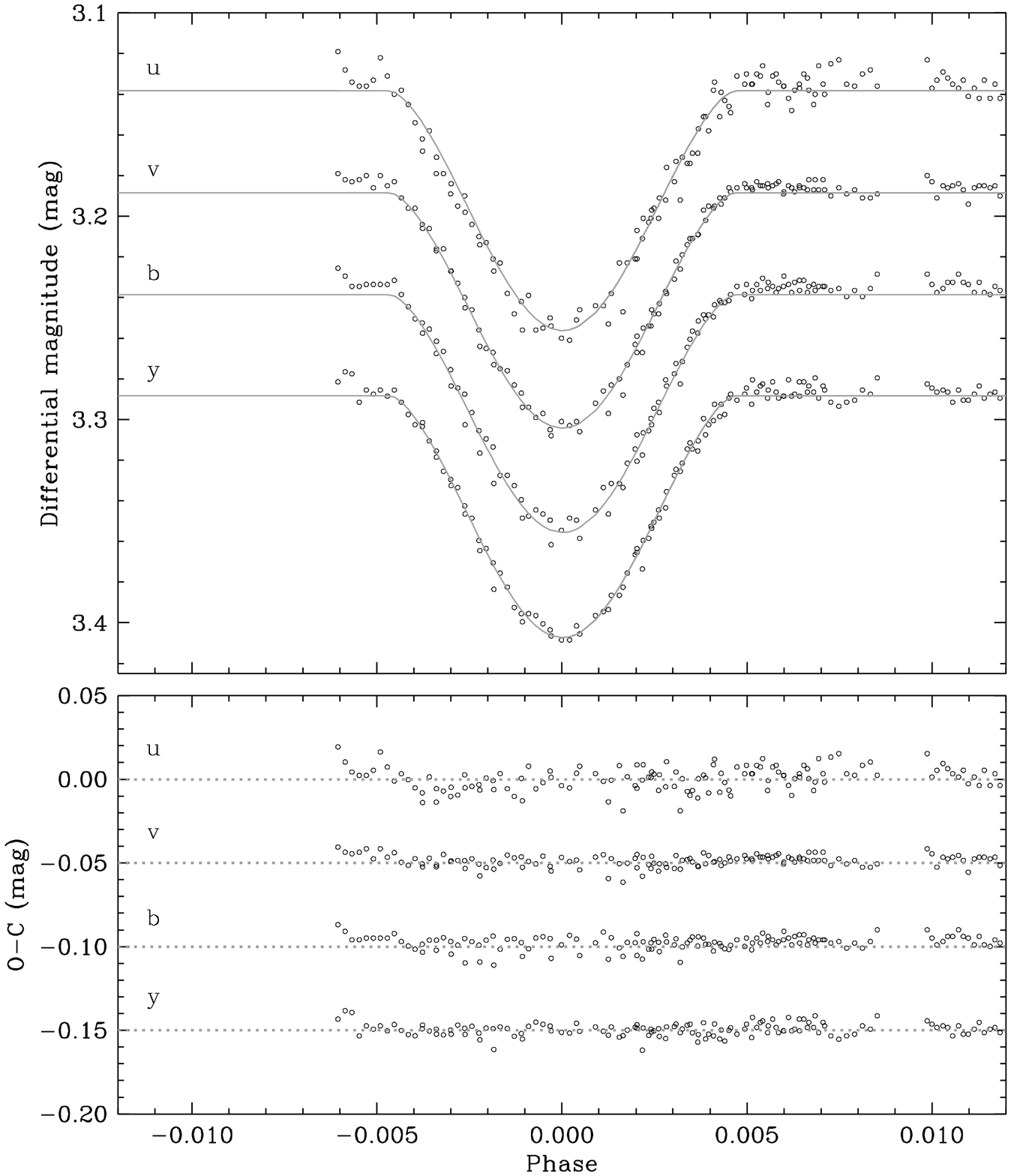}}
\caption{$uvby$ photometry for \HW\ near the primary minimum, together
with our best fit models (Table~\ref{tab:hwcma_phel}). Residuals are
shown at the bottom.}
\label{fig:hwcma_figLCprim}

\end{figure}

In previous papers of this series the LD coefficients from van Hamme
({\cite{vh93}) have generally been found to be in better agreement
than those from Claret (\cite{Claret:00}) with the values that result
when the coefficients are left free in the solutions (which cannot be
done here, for obvious reasons), and that was the case for \SW\ as
well (see Appendix~\ref{ap:LDswcma}).  Consequently, for the final
light elements of \HW\ we adopt the van Hamme (\cite{vh93})
coefficients, and take the weighted average of the $vby$
solutions (excluding $u$). These values are presented in the top portion of
Table~\ref{tab:hwcma_phel}, where the uncertainties include a
contribution from the scatter between the different bandpasses.  A
final set of solutions was carried out in each passband with the
geometric elements held fixed at these weighted average values, in
order to derive the luminosity ratios. These are listed at the bottom
of the table.  The observations and fitted curves near the primary
eclipse are shown in Fig.~\ref{fig:hwcma_figLCprim}.  The final fits
indicate only 22\% of the $y$-band light of the photometric primary
(less massive star) is blocked at phase 0.0.


\subsection{Absolute dimensions}
\label{sec:hwcma_absdim}

The mass determinations for \HW\ are both good to about 0.7\%.
Despite the difficulties described in the preceding section, the radii
are also very precise, with relative uncertainties of only 1.1\% and
1.3\% for the primary (less massive star) and secondary that we
consider realistic. These and other properties of the system are
collected in Table~\ref{tab:hwcma_absdim}.

\begin{table}[b!]
\caption{Astrophysical properties of the \HW\ system.}
\label{tab:hwcma_absdim}
\begin{minipage}{\columnwidth}
\begin{center}    
\begin{tabular}{lrr} \hline\hline    
\noalign{\smallskip}    
                     &    Primary       &    Secondary      \\ 
\noalign{\smallskip}    
\hline    
\noalign{\smallskip}    
\multicolumn{3}{l}{Absolute dimensions:}                \\ 
~~~~$M$ ($M_{\sun}$)              &$1.721 \pm 0.011$  &$1.781 \pm 0.012$ \\ 
~~~~$R$ ($R_{\sun}$)              &$1.643 \pm 0.018$  &$1.662 \pm 0.021$  \\ 
~~~~$\log g$ (cgs)                & $4.242 \pm 0.010$ & $4.247 \pm 0.011$ \\
~~~~$v \sin i$ (\kms) $^{\mathrm{a}}$   & $12.0 \pm 1.0$      & $12.0 \pm 1.0$ \\ 
~~~~$v_{\rm sync}$ (\kms) $^{\mathrm{b}}$   & $ 3.9 \pm 0.1$    & $ 4.0 \pm 0.1$  \\ 
~~~~$v_{\rm psync}$ (\kms) $^{\mathrm{c}}$  & $11.1 \pm 0.1$    & $11.2 \pm 0.2$ \\ 
~~~~$v_{\rm peri}$ (\kms) $^{\mathrm{d}}$   & $13.7 \pm 0.2$    & $13.8 \pm 0.2$  \\ 
\noalign{\smallskip}
\multicolumn{3}{l}{Radiative and other properties:} \\
~~~~$T_{\mbox{\scriptsize eff}}$ (K)       &  $7560 \pm 150$    &   $7700 \pm 150$ \\
~~~~$M_{\mbox{\scriptsize bol}}$ (mag)   &  $2.488  \pm 0.089$  &   $2.383  \pm 0.088$ \\
~~~~$\log L/L_{\sun}$ & $ 0.898 \pm 0.036$ &    $ 0.940 \pm 0.035$ \\
~~~~$BC_V$  (mag)     & $ +0.03 \pm 0.10$  &    $ +0.03 \pm 0.10$ \\
~~~~$M_V$  (mag) &    $ 2.46 \pm 0.13$     &   $ 2.35 \pm 0.13$ \\
~~~~Distance  (pc)  & \multicolumn{2}{c}{$306 \pm 15$} \\
~~~~$V-M_V$  (mag)  & \multicolumn{2}{c}{$7.43 \pm 0.10$} \\
~~~~[Fe/H] $^{\mathrm{e}}$      & $+0.33 \pm 0.10$ & $+0.28 \pm 0.10$ \\
\noalign{\smallskip}
\multicolumn{3}{l}{Photometric indices:}    \\
~~~~$V$ $^{\mathrm{f}}$ &         $ 9.996 \pm 0.012$  &        $ 9.892 \pm 0.011$\\              
~~~~$(b-y)$ $^{\mathrm{f}}$ &     $ 0.139 \pm 0.005$  &        $ 0.118 \pm 0.005$\\
~~~~$m_1$ $^{\mathrm{f}}$   &     $ 0.225 \pm 0.011$  &        $ 0.231 \pm 0.011$\\
~~~~$c_1$ $^{\mathrm{f}}$   &     $ 0.817 \pm 0.016$  &        $ 0.858 \pm 0.016$\\
~~~~$E(b-y)$    & \multicolumn{2}{c}   {$0.026 \pm 0.010$} \\
\noalign{\smallskip}            
\hline
\noalign{\smallskip}
\end{tabular}            
\begin{list}{}{}
\item[$^{\mathrm{a}}$] Measured projected rotational velocity.
\item[$^{\mathrm{b}}$] $v \sin i$ expected for synchronous rotation.
\item[$^{\mathrm{c}}$] $v \sin i$ expected for pseudo-synchronous rotation.
\item[$^{\mathrm{d}}$] Expected $v \sin i$ if synchronized at periastron.
\item[$^{\mathrm{e}}$] Values representative of the surface layers only.
\item[$^{\mathrm{f}}$] Not corrected for interstellar absorption/reddening.
\end{list}
\end{center}
\textsc{Note:} The bolometric corrections ($BC_V$) are adopted from
Flower (\cite{flower96}), along with with $T_{\rm eff\sun} = 5780$ K,
and $M_{bol\sun} = 4.73$ (see Torres \cite{Torres:10b}). An additional
error contribution of 0.10~mag is added in quadrature to the $BC_V$
uncertainty propagated from the temperature uncertainties.
\end{minipage}
\end{table}                                  

With the $uvby\beta$ indices out of eclipse from Clausen et al.\
(\cite{Clausen:08}) and the calibration by Crawford
(\cite{Crawford:79}) we derive an estimate of the interstellar
reddening toward \HW\ of $E(b-y) = 0.026$. We assign to this a
conservative uncertainty of 0.01 mag. The corresponding extinction is
$A(V) = 0.11 \pm 0.04$. A rather similar value is obtained from the
dust maps of Hakkila et al.\ (\cite{Hakkila:97}), which give $A(V) =
0.13$ or $E(b-y) = 0.031$ after correction for the distance to the
binary.  On the other hand, Schlegel et al.\ (\cite{Schlegel:98})
suggest larger values of $A(V) = 0.20$ and $E(b-y) = 0.049$. In the
following we adopt the more reliable photometric estimate.
Table~\ref{tab:hwcma_absdim} includes also the individual Str\"omgren
indices of the components, derived from the combined $uvby$ magnitudes
and the light ratios used in our photometric solutions. After
de-reddening, these lead to photometric temperatures of $7777 \pm
102$\,K and $7972 \pm 107$\,K using the calibration of Napiwotzki et
al.\ (\cite{nsw93}), in which the uncertainties include observational
errors but not the unknown dispersion of the calibration. These
estimates are 200--250\,K hotter than the adopted spectroscopic
values.

In addition to the two consistent estimates of the effective
temperatures reported earlier from the CfA/DS and FIES spectra, we
obtained another measure by disentangling the DS spectra and using a
cross-correlation procedure similar to that described in
Sect.~\ref{sec:swcma_spec_orb} for \SW, but adapted for single-lined
spectra. The disentangling was carried out with a revised version
provided by E.\ Sturm of the original procedure introduced by Simon \&
Sturm (\cite{Simon:94}).  The results are 7500\,K and 7620\,K for the
primary and secondary.\footnote{While the cross-correlation
technique applied to these disentangled spectra is able to extract a
useful estimate of $T_{\rm eff}$ because it makes use of the full
spectrum at once, there are not enough isolated lines of iron and
other elements in this narrow wavelength region to attempt a detailed
abundance analysis as was done earlier with the FIES spectrum.}  A
further spectroscopic estimate of the temperatures was made from
three spectra of \HW\ collected with the TRES instrument (Szentgyorgyi
\& F\H{u}r\'esz \cite{Szentgyorgyi:07}) on the 1.5\,m reflector at the
F.\ L.\ Whipple Observatory ($\lambda/\Delta\lambda \approx 44\,000$,
$\lambda\lambda$3900--8900\,\AA). These observations were initially
intended to support the abundance determinations in
Sect.~\ref{sec:hwcma_abund}, but turned out to be too weak for that
purpose. Application of the TODCOR-based cross-correlation procedures
used before for composite spectra gave $T_{\rm eff}$ values of 7610\,K
and 7670\,K for the primary and secondary, and a light ratio of
$L_p/L_s = 0.89 \pm 0.02$ at a mean wavelength of 5187\,\AA.  Finally,
we applied the same TODCOR analysis technique to the FIES spectrum we
used earlier for the abundance analysis, and obtained 7570\,K and
7740\,K along with an identical light ratio of $0.89 \pm 0.02$. The
above determinations are all in good agreement, and establish that the
primary (less massive star) is slightly but significantly cooler than
the secondary.  We adopt a straight average of the five temperature
estimates, 7560\,K and 7700\,K for the primary and secondary, to which
we attach an uncertainty of 150\,K. The corresponding spectral type is
approximately A6. The average of the three independent spectroscopic
light ratio determinations is $L_p/L_s = 0.90 \pm 0.02$.

The luminosities and distance were derived in the same way as for \SW.
The system distance is $306 \pm 15$\,pc. Nearly identical values of
$307 \pm 20$\,pc and $306 \pm 20$\,pc obtained separately for the
primary and secondary indicate excellent internal consistency.  \HW\
and \SW\ are only 2\farcm5 apart on the sky and \HW\ is only half as
distant as \SW, yet the reddening we find for \HW\ is slightly
\emph{larger}.  This may indicate that interstellar extinction is
patchy in this direction.

The measured projected rotational velocities for both components of
\HW\ are consistent with the pseudo-synchronous velocities, assuming
co-aligned orbital and spin axes (see Table~\ref{tab:hwcma_absdim}).


\section{Discussion}
\label{sec:discussion}

The following sections present a comparison of the measured properties
for \SW\ and \HW\ against predictions from the theory of stellar
evolution, stellar structure, and tidal evolution.


\subsection{Stellar evolution}
\label{sec:stellarevolution}

Accurate mass, radius, and effective temperature determinations,
especially when accompanied by a measurement of the chemical
composition, are among the most powerful constraints afforded by
eclipsing binary stars to test models of stellar structure and
evolution (see, e.g., Andersen \cite{ja91}, Torres et al.\
\cite{Torres:10a}, and references therein). When these four quantities
are available for both components in the system, there are essentially
no free parameters in the comparison with publicly available models:
to the extent that the measurements are accurate, the models will
either be successful in reproducing all observations simultaneously
within their errors, or they will fail.\footnote{Additional variables
such as the helium abundance, mixing length parameter, convective core
overshooting, etc., have generally already been chosen in advance by
the modelers in published tables, and are not ``tunable'' by the
user. Specific models for a particular system can of course be
computed with different values of these additional variables, as we do
in one example below, adding more freedom to the fits.}

In the case of Am stars such as \SW\ and \HW, however, the measured
abundances do not represent the bulk composition of the stars, which
is the relevant quantity when comparing with models. Instead, they are
only a reflection of local changes in the surface layers attributed to
diffusion processes, which give the spectra of these objects their
peculiar appearance. Metallicity therefore remains a free parameter
when testing models for these stars. Nevertheless, because detailed
abundance patterns in Am stars vary from case to case and are not yet
completely understood, a comparison with models is still of
considerable interest -- even if less constraining -- to establish the
precise evolutionary state of the stars. Except for second-order
effects due to limb darkening, absolute masses and radii are
essentially independent of the theory being tested. For \HW, however,
our light-curve solutions have had to make additional use of model
atmospheres to constrain the geometry of the system, because of the
lack of secondary eclipses. We point out, though, that those models
were used only in a differential sense (to determine \emph{ratios} of
central surface fluxes or bandpass-specific luminosities), so in
effect the dependence on theory is very weak.

\begin{figure}[t]
\epsfxsize=85mm
\epsfbox{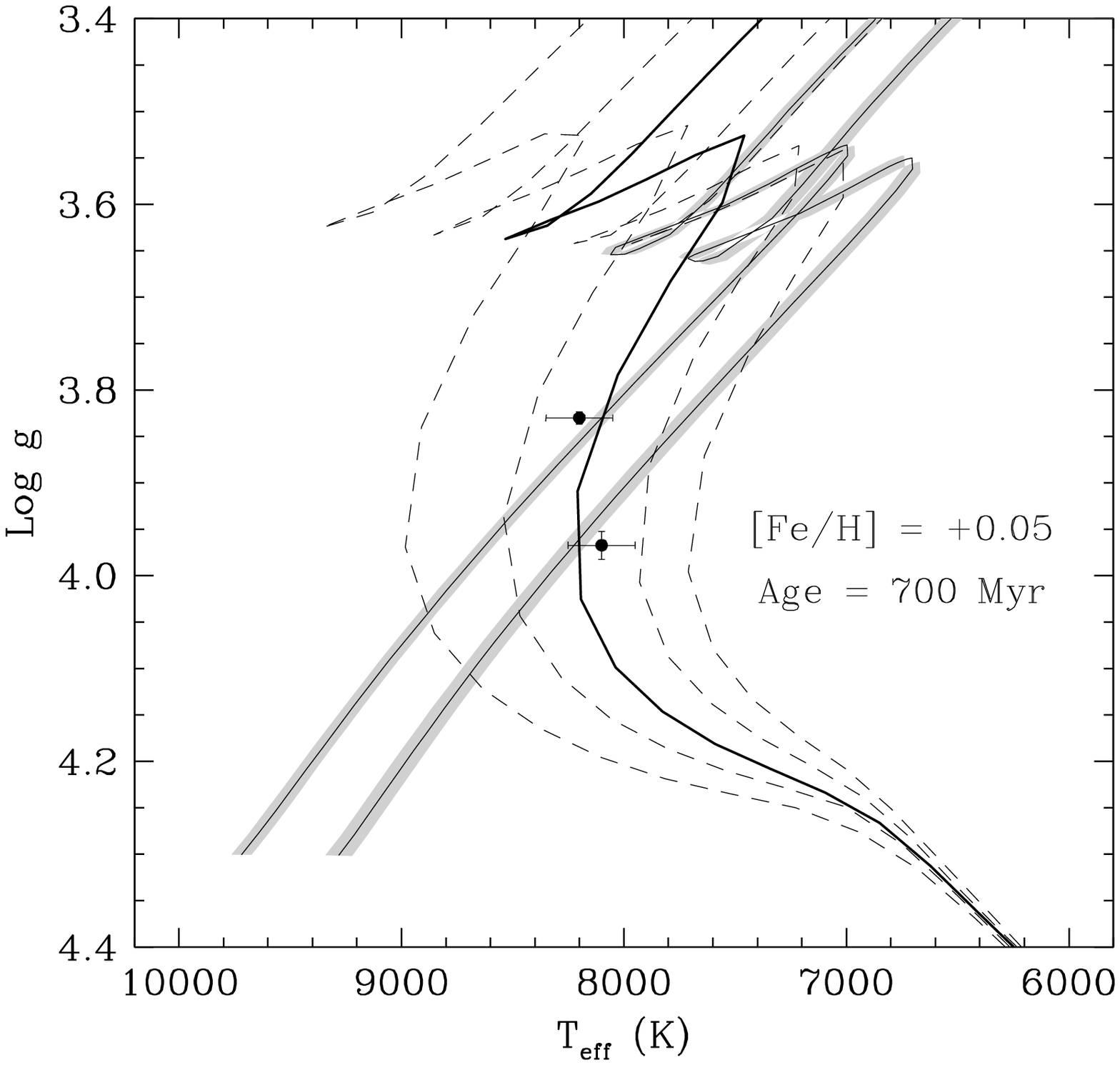}
\caption{\label{fig:swcma_tg} Measurements for \SW\ compared against
Yonsei-Yale models by Yi et al.\ (\cite{Yi:01}) for [Fe/H] $= +0.05$
and [$\alpha$/Fe] = 0.0.  Evolutionary tracks for the measured masses
are shown with solid lines and shaded areas indicating the uncertainty
in the location of each track coming from the mass errors. Isochrones
from 500~Myr to 900~Myr in steps of 100~Myr are represented by the
dashed lines. The best fitting 700~Myr isochrone is drawn with a thicker
line.}
\end{figure}

\begin{figure}
\epsfxsize=85mm
\epsfbox{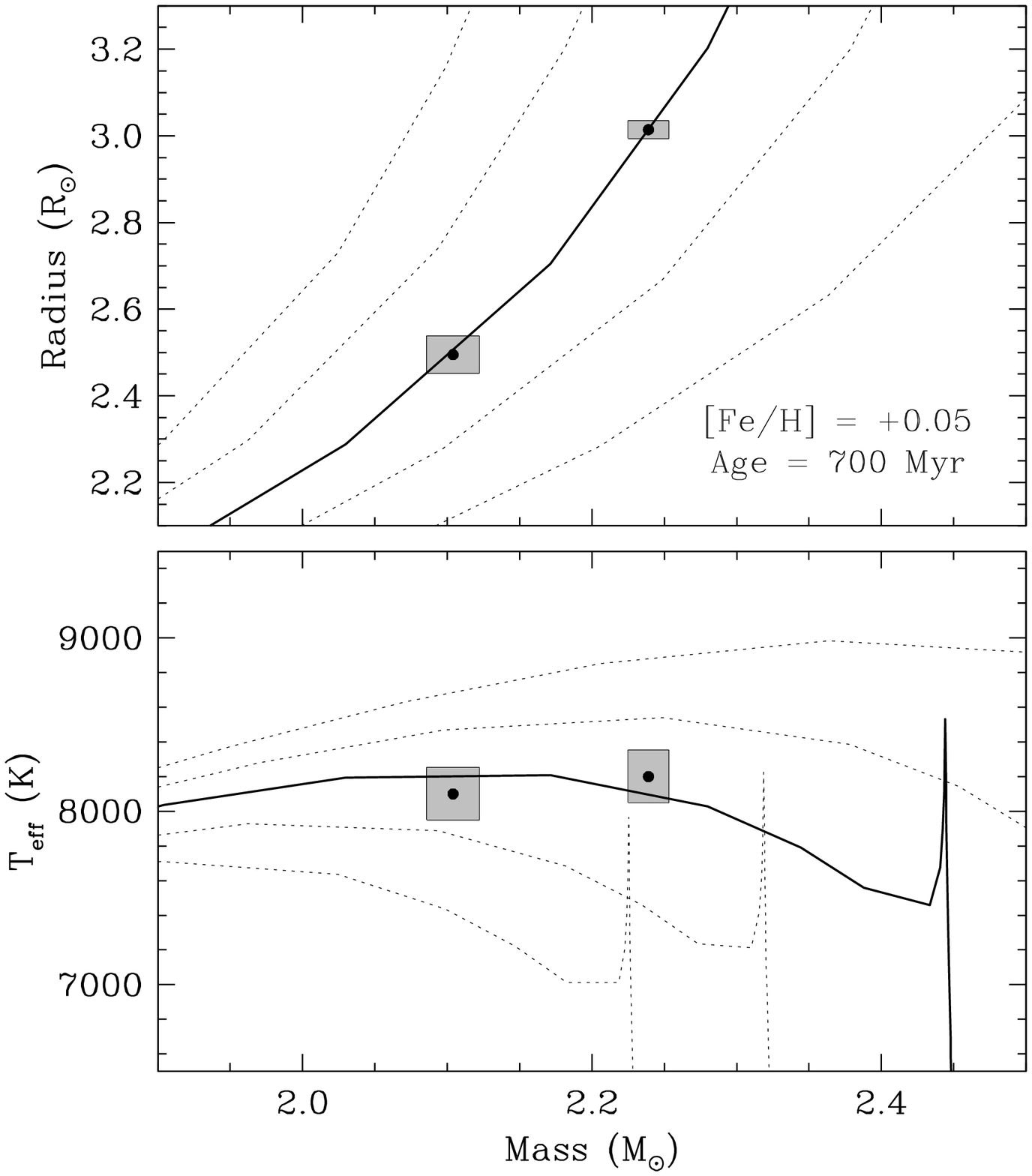}
\caption{\label{fig:swcma_mr} Effective temperatures and radii for
\SW\ shown as a function of the measured mass, with the observational
errors represented by the shaded boxes. They are compared against
[Fe/H] $= +0.05$ Yonsei-Yale isochrones by Yi et al.\ (\cite{Yi:01})
from 500~Myr to 900~Myr, in steps of 100~Myr, with [$\alpha$/Fe] =
0.0. The best-fitting model is indicated with a heavier line.}
\end{figure}

In this section we compare the absolute dimensions of \SW\ and \HW\
with stellar evolution models from the Yonsei-Yale series (Yi et al.\
\cite{Yi:01}; Demarque et al.\ \cite{Demarque:04}), which have been
shown previously to match observations very well. We have chosen these
models primarily because they are provided with software to
interpolate evolutionary tracks (and isochrones) for any given mass
and metallicity, the mass being among the most accurately known
properties for these stars.

The $\log g$ vs.\ $T_{\rm eff}$ diagram for \SW\ in
Fig.~\ref{fig:swcma_tg} allows a simultaneous comparison of the
models against the three key observables: $M$, $R$, and $T_{\rm eff}$.
Evolutionary tracks corresponding to a composition slightly above
solar ([Fe/H] $= +0.05$, with no $\alpha$-element enhancement) are
shown for the exact masses we measure (Table~\ref{tab:swcma_absdim}),
with the gray areas representing the uncertainty in the location of
the tracks that comes from the mass errors (in this case, a small
fraction of the mass difference). Isochrones are also displayed for
ages 500--900~Myr, in steps of 100~Myr. The models for this
composition provide a very good fit to the observations, within the
uncertainties, and the age inferred is approximately 700~Myr
(indicated with the thicker isochrone). \SW\ is seen to be somewhat
evolved, and currently near the mid-point of its main sequence phase.
The implied bulk composition inferred from theory, [Fe/H] $= +0.05$,
is $\sim$0.4--0.5~dex lower than the iron abundance measured at the
surface ([Fe/H] $= +0.49$/+0.61 for the primary and secondary,
respectively), which may be interpreted as giving a rough measure of
the \emph{true} enhancement of the iron-peak elements in the
photospheres of these stars.
Strictly speaking, however, this factor of three enhancement may only
be a lower limit, as the effects of metallicity and $\alpha$-element
enhancement in the models can trade off against each other to some
extent, leading to mass tracks and isochrones that fit the
observations equally well.  For example, a match as good as seen above
is achieved also with [Fe/H] $= -0.10$ and [$\alpha$/Fe] $= +0.22$,
with [Fe/H] $= -0.20$ and [$\alpha$/Fe] $= +0.36$, etc.

Separate diagrams of the radius and effective temperature as a
function of mass are seen in Fig.~\ref{fig:swcma_mr}, along with the
same isochrones from the previous figure. The models provide a
virtually exact match to the mass and radius of the components at
700~Myr, and a good fit to the temperatures as well.

\begin{figure}[t]
\epsfxsize=85mm
\epsfbox{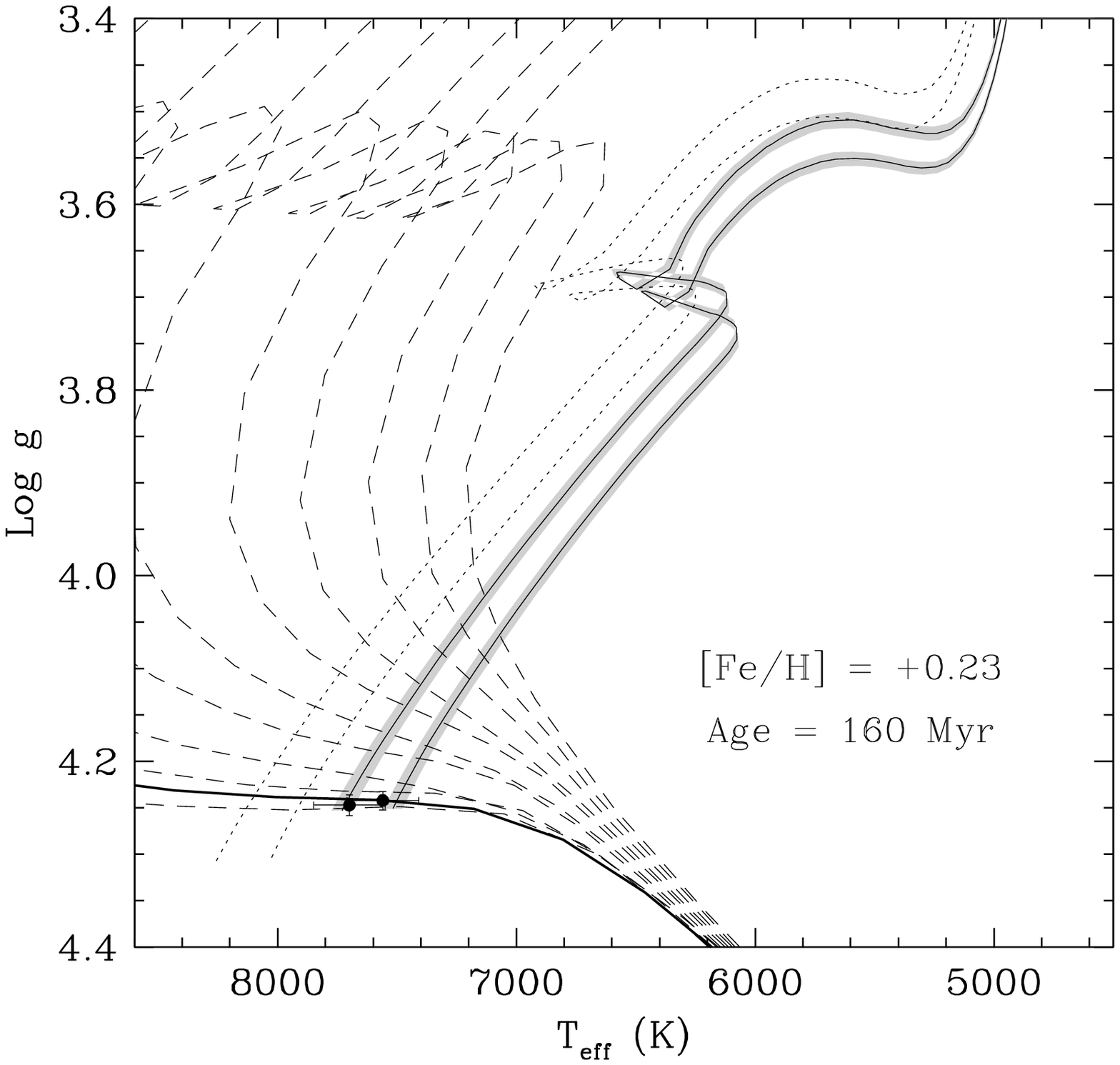}
\caption{\label{fig:hwcma_tg} Similar to Fig.~\ref{fig:swcma_tg}, but
for \HW. The best-fitting metallicity is [Fe/H] $= +0.23$. Solar
metallicity tracks are shown for reference (dotted lines). The dashed
lines represent Yonsei-Yale isochrones from 100~Myr to 1~Gyr in steps
of 100~Myr. The best fitting 160~Myr isochrone is drawn with a thick
solid line.}
\end{figure}


\begin{figure}[t]
\epsfxsize=85mm
\epsfbox{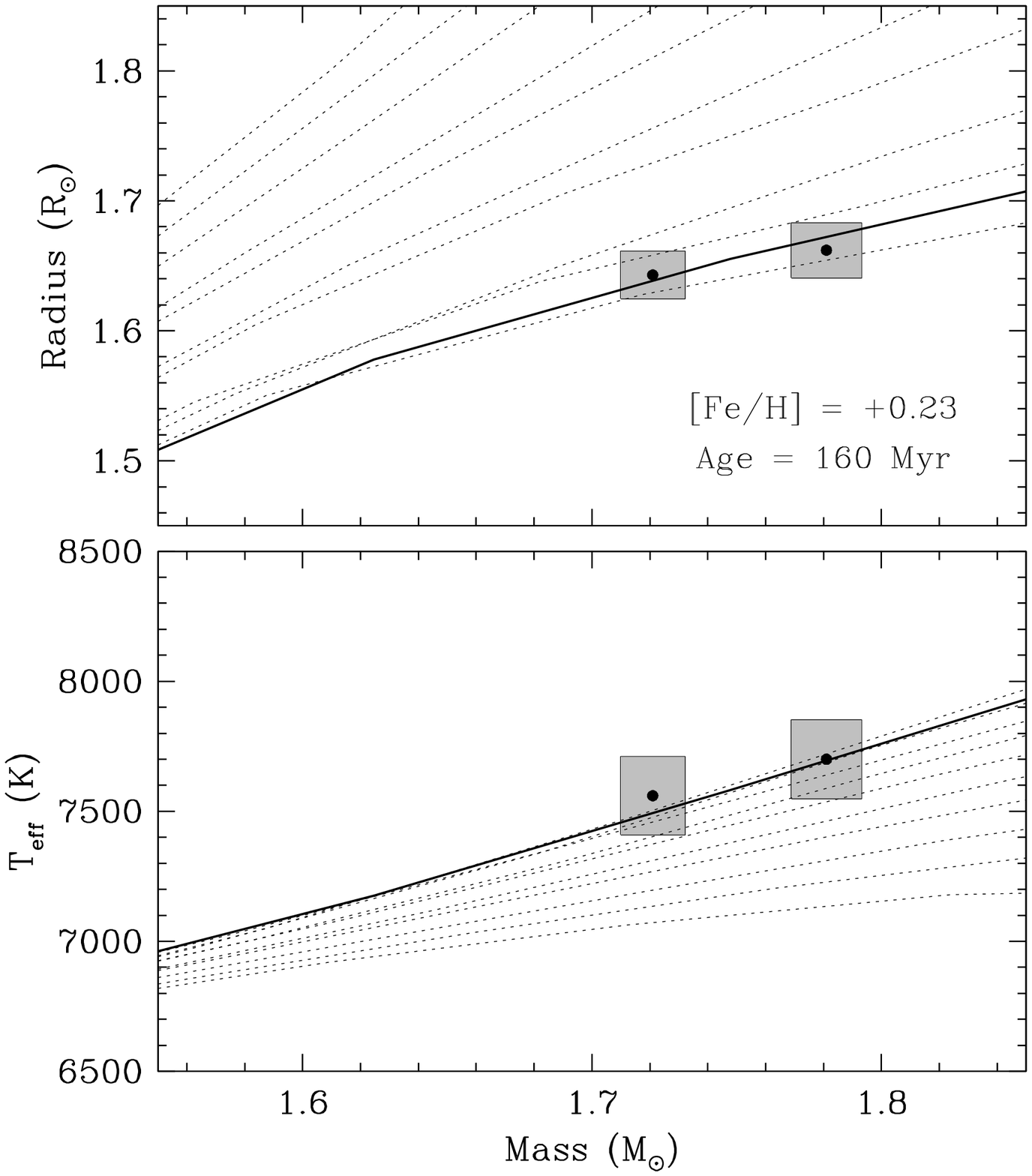}
\caption{\label{fig:hwcma_mr} Similar to Fig.~\ref{fig:swcma_mr}, but
for \HW. In this case the best-fitting metallicity is [Fe/H] $=
+0.23$. The dotted lines represent Yonsei-Yale isochrones from 100~Myr
to 1~Gyr in steps of 100~Myr. The best fitting 160~Myr model is drawn
with a heavier solid line.}
\end{figure}

Similar diagrams for \HW\ can be seen in Fig.~\ref{fig:hwcma_tg} and
Fig.~\ref{fig:hwcma_mr}. In this case, the model composition that
best fits the observations is [Fe/H] $= +0.23$ (with [$\alpha$/Fe] =
0.0), and the implied age is approximately 160~Myr. Solar composition
tracks are shown for reference as dotted lines in
Fig.~\ref{fig:hwcma_tg}. \HW\ is essentially on the zero-age main
sequence. To the extent that [Fe/H] $= +0.23$ represents the overall
composition of this system, it would appear that the true enhancement
of the iron-peak elements in the surface layers is a modest 0.1~dex in
this case, since the measured photospheric abundances are [Fe/H] $=
+0.33$/+0.32.
As before, this conclusion is dependent on the assumed [$\alpha$/Fe]
enhancement, which is inaccessible to observation in both of these
systems.  The higher-than-solar bulk [Fe/H] $= +0.23$ composition for
\HW\ suggested by the models is somewhat uncommon for stars in the
solar neighborhood.  If one enforces in the models a more typical
solar-like iron abundance, the observations are about as well
reproduced by increasing the [$\alpha$/Fe] ratio to +0.40. This
combination would imply a true surface-layer iron enhancement due to
the Am phenomenon in \HW\ of $\sim$0.3~dex, and the inferred age would
be the same as before.  However, a value of [$\alpha$/Fe] as large as
this is on the high side of the observed distribution for stars with
solar iron abundance, which makes us somewhat skeptical of this
scenario.
Another possible way of reproducing the \HW\ observations with an iron
composition closer to solar is to alter the helium abundance.  Tests
with the Granada models of Claret (\cite{Claret:04}) indicate,
however, that the radii and effective temperatures can only be matched
simultaneously with a very low helium abundance of $Y = 0.24$, at ages
of 440--500\,Myr that are significantly older than before.  A helium
abundance this close to the primordial value seems unlikely.

Comparisons (not shown) of the observations for \HW\ and \SW\ with the
standard ([$\alpha$/Fe] = 0.0) Victoria-Regina stellar evolution
models of VandenBerg et al.\ (\cite{vr06}) and with the Granada models
of Claret (\cite{Claret:04}) give results that are quite consistent
with those from the Yonsei-Yale models, indicating similar overall
abundances and evolutionary ages.


\subsection{Internal structure}
\label{sec:internalstructure}

The observed apsidal motion for \SW, $\dot\omega = 0.00067 \pm
0.00021$~deg~cycle$^{-1}$ (Clausen et al.\ \cite{Clausen:08}), is a
measure of the average degree of central mass concentration of the
components. The corresponding average apsidal motion constant for the
system is $\log \bar k_{2,\rm obs} = -2.57 \pm 0.30$. Using the
absolute dimensions of the stars from Table~\ref{tab:swcma_absdim},
the models by Claret (\cite{Claret:04}) predict a theoretical value of
$\log \bar k_{2,\rm theo} = -2.582 \pm 0.050$, in which the formal
uncertainty does not account for possible systematic errors in the
models. While this is consistent with the empirical value within the
errors, the close agreement may be accidental given the difficulty of
the measurement (the apsidal period is long: $U = 14\,900 \pm
4700$\,yr). The General relativistic contribution ($\dot\omega_{\rm
  GR} = 0.00034$~deg~cycle$^{-1}$) amounts to about 50\% of the total
apsidal motion.

While the eccentric system \HW\ is also expected to display apsidal
motion, the lack of secondary minima precludes such a measurement
based on eclipse timings. Spectroscopic detection of the effect would
be quite challenging. The models of Claret (\cite{Claret:04}) predict
a value of $\dot\omega = 0.00023$~deg~cycle$^{-1}$, and a
corresponding apsidal motion constant of $\log \bar k_{2,\rm theo} =
-2.410 \pm 0.050$. The estimated period of this effect is $U =
88\,900$\,yr, six times longer than in \SW. In the case of \HW\ the
relativistic contribution completely dominates, accounting for 97\% of
the total apsidal motion.


\subsection{Tidal evolution}
\label{sec:tidalevolution}

In addition to their binarity, Am stars typically display slow
rotation compared to field stars of the same spectral type. The
components of \SW\ and \HW\ are no exception.  We noted earlier that
while the primary of \SW\ appears to be pseudo-synchronized with the
orbital motion, the secondary is not, and rotates at only half the
corresponding speed. This assumes the spin axes are parallel to the
axis of the orbit, a condition that is almost universally taken for
granted in binaries. Here we investigate whether synchronization is
expected at all in these systems according to tidal theory, and if so,
whether the situation for the secondary of \SW\ might be understood in
terms of spin-orbit misalignment. We examine also the predicted
evolution of the orbital eccentricity.

\begin{figure*}
\epsfxsize=175mm
\hskip 0.07in\epsfbox{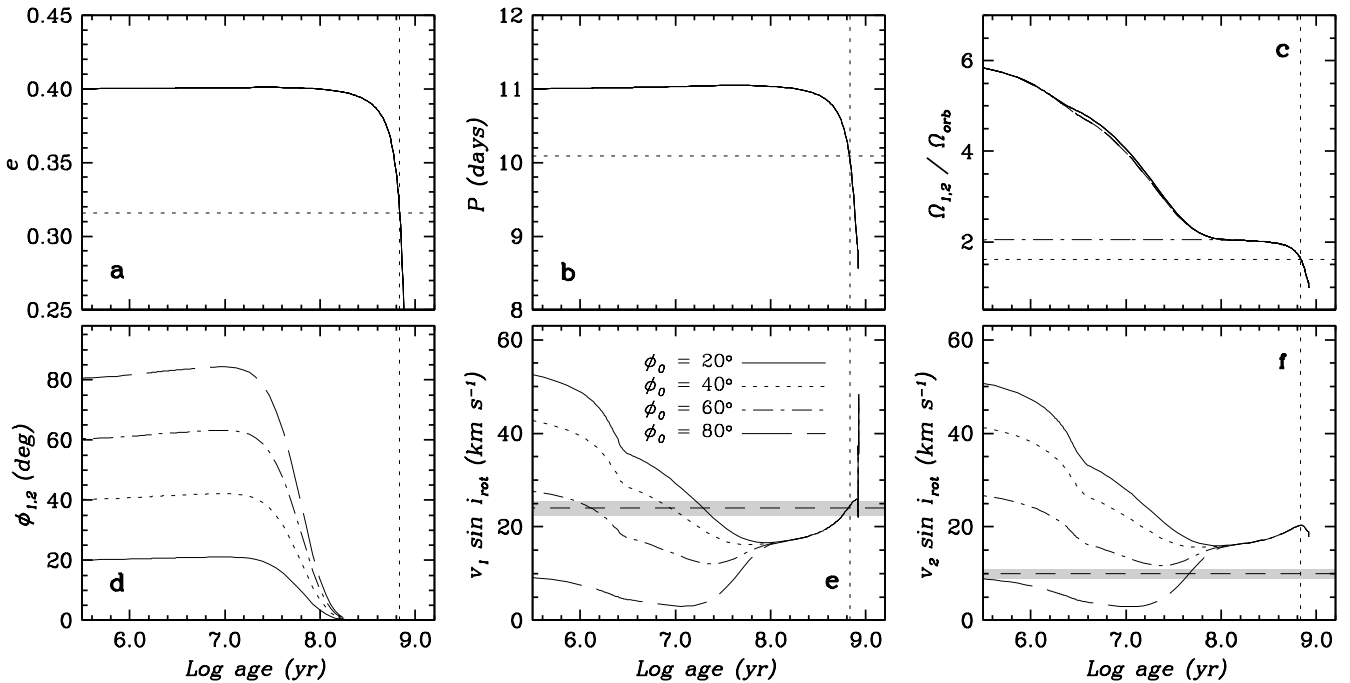}
\caption{Tidal evolution calculations for \SW, following the
prescription by Hut (\cite{Hut:81}). The vertical dotted line in each
panel represents the current evolutionary age of 680\,Myr ($\log\tau =
8.833$), based on the models by Claret (\cite{Claret:04}). (a)
Eccentricity as a function of time; (b) Orbital period; (c) Normalized
angular rotation rate of each star. The dot-dashed line represents the
evolution of the pseudo-synchronous value, and the horizontal dotted
line is the pseudo-synchronous rate at the present age. (d) Angle
between the equator of each star and the plane of the orbit. The four
curves correspond to different initial values as given in the next
panel. (e) Theoretical projected rotational velocity of the primary as
a function of time, for four different initial values of the
spin-orbit angle, $\phi_0$. The measured $v \sin i_{\rm rot}$ and its
uncertainty are indicated with the horizontal dashed line and shaded
area. Theory agrees with observation at the current age of the
system. (f) Same as (e), for the secondary. In this case theory is not
able to match the observation, regardless of the initial values
assumed for $\Omega/\Omega_{\rm orb}$ and $\phi$.\label{fig:tidal_sw}}
\end{figure*}

We have integrated the differential equations given by Hut
(\cite{Hut:81}) for the time evolution of the semimajor axis
($da/dt$), the eccentricity ($de/dt$), the angular rotation rates of
both components ($d\Omega_1/dt$, $d\Omega_2/dt$), and the angle
between the plane of the orbit and the equator of each star
($d\phi_1/dt$, $d\phi_2/dt$).  These six, coupled equations were
integrated simultaneously using a 4th-order Runge-Kutta method, with
the stellar properties interpolated at each time step from appropriate
evolutionary tracks from Claret (\cite{Claret:04}). The turbulent
dissipation timescale for the late evolutionary phases with convective
envelopes was taken to be $(M R^2/L)^{1/3}$, where $M$, $R$, and $L$
are the mass, radius, and luminosity of the star. For phases in which
the envelopes are radiative (the most relevant here) the timescales
adopted follow closely those in Eq.\,17 and Eq.\,18 by Claret \&
Cunha (\cite{Claret:97}).

The initial conditions are of course not known, so they are in effect
free parameters of the model. For the orbital period and eccentricity
we have adjusted the starting values so as to match $P$ and $e$ at the
current evolutionary age of each system. Their time evolution for \SW\
is shown in Fig.~\ref{fig:tidal_sw}a and Fig.~\ref{fig:tidal_sw}b,
in which the vertical dotted lines mark the current age of about
680\,Myr ($\log\tau = 8.833$), according to the Granada models
employed in this section. Based on these calculations the orbit is
expected to circularize at an age of $\tau = 835$\,Myr ($\log\tau =
8.921$) for this system.

The evolution of the angular rotation rate $\Omega = 2\pi/P_{\rm rot}$
of each star in \SW\ is seen in Fig.~\ref{fig:tidal_sw}c, where for
convenience we have normalized it to the orbital rate, $\Omega_{\rm
orb} = 2\pi/P_{\rm orb}$.  We have no observational constraint on the
rotation rates, so the initial values were arbitrarily set to
$\Omega_1/\Omega_{\rm orb} = \Omega_2/\Omega_{\rm orb} = 6.0$,
reasonable for A-type stars that usually rotate very rapidly (at least
in the field). A dot-dashed line in this figure indicates the
pseudo-synchronous rate at each age, and the horizontal dotted line is
the current pseudo-synchronous rate.  These calculations suggest that
pseudo-synchronization for both stars in this system was reached
early-on, at an age of roughly 100\,Myr ($\log\tau = 8.0$). This would
indeed appear to be the case for the primary of \SW, based on its
measured $v \sin i$ value, but not for the secondary.

The degree of spin-orbit alignment is represented by the angle $\phi$,
but once again, we have no constraint on this quantity. Furthermore,
the relation between $\phi$ and the orbital and rotational inclination
angles $i_{\rm orb}$ and $i_{\rm rot}$, both measured with respect to
the line of sight, is given by
\begin{equation}
\cos\phi = \cos i_{\rm orb} \cos i_{\rm rot} + \sin i_{\rm orb} \sin
i_{\rm rot} \cos\lambda~,
\end{equation}
which involves an unknown angle $\lambda$ between the sky-projected
angular momentum vectors of the orbit and the stellar spin.

\begin{figure*}[t]
\epsfxsize=175mm
\hskip 0.07in\epsfbox{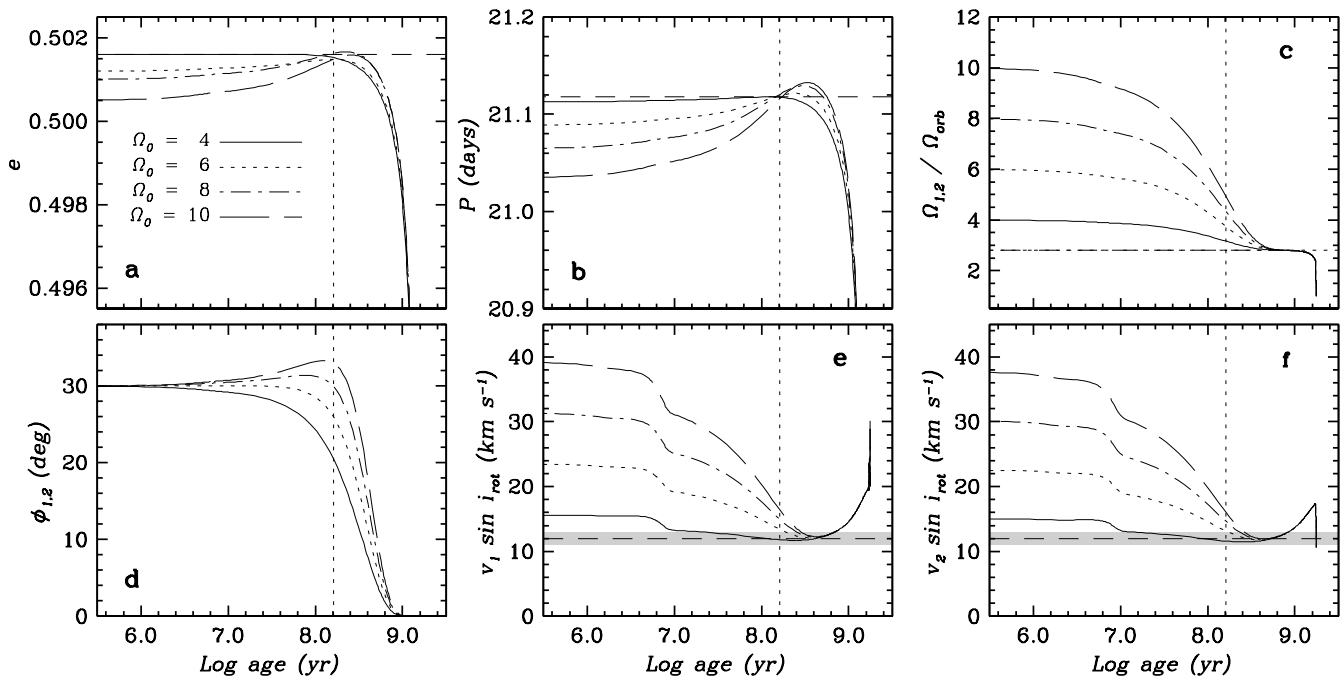}
\caption{Similar to Fig.~\ref{fig:tidal_sw}, but for the \HW\
system. In this case the results are illustrated for four initial
values of $\Omega/\Omega_{\rm orb}$ ($\Omega_0$ for short) as labeled
in panel (a), and a single initial value for $\phi$ of 30\degr. With
the boundary conditions tuned in this way, theory matches the measured
rotational velocities of both components shown in (e) and (f). Neither
star is predicted to be synchronized, and the calculations suggest
their spin axes have not yet been aligned with the axis of the orbit.
\label{fig:tidal_hw}}
\end{figure*}

The spectroscopically measured projected rotational velocities of the
stars, which we refer to more properly now as $v \sin i_{\rm rot}$, do
provide an indirect constraint on a combination of theoretically
predictable quantities, but this still involves the unknown angle
$\lambda$. Given that $i_{\rm orb}$ is rather close to 90\degr\ in
both \SW\ and \HW\, we may make the approximation that $\cos\phi
\approx \sin i_{\rm rot} \cos\lambda$. We then have
\begin{equation}
\label{eq:tidal}
v \sin i_{\rm rot} \approx {2\pi\over P_{\rm orb}} {\Omega\over
\Omega_{\rm orb}} {\cos\phi\over \cos\lambda} R~.
\end{equation}
All quantities on the right-hand side of this equation are either
known from stellar evolution calculations ($R$), or can be computed
from the solution of the differential equations for tidal evolution,
with the exception of the angle $\lambda$, which depends on the
observer's viewpoint. In order to make progress, we ignore this term
for the moment (or equivalently, we consider $\lambda$ to be small),
so that $\cos\phi \approx \sin i_{\rm rot}$. 

The evolution of the alignment angle $\phi$ for \SW\ is displayed in
Fig.~\ref{fig:tidal_sw}d for four different initial values (20\degr,
40\degr, 60\degr, 80\degr). The curves for the primary and secondary
are nearly indistinguishable, so only those for the primary are shown.
As seen from the convergence of the curves towards zero, spin-orbit
alignment for both stars in \SW\ is reached at the age of about
200\,Myr ($\log\tau \approx 8.3$), which is less than a third of the
current age of the system.  The predicted evolution of $v \sin i_{\rm
rot}$ for each star, computed from Eq.\,\ref{eq:tidal}, is shown in
Fig.~\ref{fig:tidal_sw}e and Fig.~\ref{fig:tidal_sw}f compared against
the measured values (dashed line and shaded uncertainty region).

Independently of the initial values of $\phi$ and $\Omega/\Omega_{\rm
orb}$, the predictions from tidal evolution theory cannot be made to
match the anomalously slow rotation rate of the secondary of \SW.  The
angle $\lambda$ in Eq.\,\ref{eq:tidal} that we have previously
ignored is inconsequential for this discussion, as any value greater
than zero only makes the disagreement worse.  Evidently current tidal
theory is incomplete.  An effect not considered here is the possible
decoupling between the core of the star and the external layers, which
suffer more directly the actions of tidal forces. Since the
spectroscopically measured $v \sin i_{\rm rot}$ corresponds only to
the layers that we see, it is quite possible that the larger rotation
rate predicted by theory is more representative of a more rapidly
rotating core than of the outer envelope, or at the very least is some
average of the two, which would be larger than the measured value.

Similar tidal calculations for \HW\ are shown in
Fig.~\ref{fig:tidal_hw}. In this case we have chosen to illustrate
the effects of a change in the initial values of $\Omega/\Omega_{\rm
orb}$, rather than $\phi$. As before, we have adjusted the boundary
conditions for the period and eccentricity to match the measured
values at the current age, which is 160\,Myr ($\log\tau = 8.21$)
according to the Claret (\cite{Claret:04}) models. As the evolution of
both of these quantities (Fig.~\ref{fig:tidal_hw}a and
Fig.~\ref{fig:tidal_hw}b) depends on $\Omega$ (see Hut
\cite{Hut:81}), the starting values for $P$ and $e$ are different in
each case. The calculations indicate that orbital circularization will
occur at an age of $\sim$1.75\,Gyr ($\log\tau = 9.244$). 

The time dependence of $\Omega/\Omega_{\rm orb}$ appears in
Fig.~\ref{fig:tidal_hw}c, and influences the late evolution of the
angle $\phi$ between the equator and orbital plane, as seen in
Fig.~\ref{fig:tidal_hw}d. These two panels indicate that neither
star is expected to be pseudo-synchronized at the current age, nor has
spin-orbit alignment been achieved. Pseudo-synchronization is
predicted to happen at $\sim$600\,Myr ($\log\tau = 8.78$), and
alignment at an age of approximately 1\,Gyr. The fact that we do see
agreement within the measurement errors between the observed
rotational velocities and the predicted pseudo-synchronous velocities
at the current epoch as listed in Table~\ref{tab:hwcma_absdim}
suggests that this may be due to a combination of non-synchronized
angular rotation rates ($\Omega_{1,2}$) and slightly misaligned spin
axes ($\phi_{1,2} \approx 20\degr$ at the current epoch for our
particular choice of the initial values for $\Omega_{1,2}$; see
Fig.~\ref{fig:tidal_hw}d).  The theoretical projected rotational
velocities are plotted in Fig.~\ref{fig:tidal_hw}e and
Fig.~\ref{fig:tidal_hw}f. The lower curves match the measured $v
\sin i_{\rm rot}$ of each star at the current age, and correspond to
initial $\Omega/\Omega_{\rm orb}$ values of 4.0 for both stars, and
initial angles $\phi_{1,2} = 30$\degr\ tuned to produce the agreement.

The asynchronous rotation of the secondary of \SW\ is not unusual for
Am stars, or for normal A stars for that matter. The compilation of
Torres et al.\ (\cite{Torres:10a}) includes several such systems with
well determined properties that show the same lack of synchronization.
An extreme case is V459~Cas (Lacy \cite{Lacy:04}), a pair of A1m stars
both rotating 3.5--4.5 times faster than pseudo-synchronous.  There is
no obvious correlation that we can see between asynchronous rotation
and other global properties of Am or normal A stars. This issue has
also been studied by others (Budaj \cite{Budaj:96}, \cite{Budaj:97};
Iliev et al.\ \cite{Iliev:06}; Prieur et al.\ \cite{Prieur:06}), but
no credible pattern has emerged.


\section{Final remarks}

\SW\ and \HW\ join the very small ranks of Am stars with
accurately known absolute dimensions (better than 2\% relative errors
in $M$ and $R$) that also have their detailed abundances
determined, in this case for a dozen or more elements in each
star. We find that the measured masses, radii, and temperatures are
well matched by stellar evolution models such as those by Yi et al.\
(\cite{Yi:01}), with inferred bulk abundances of [Fe/H] $= +0.05$ and
+0.23 and ages of $\sim$700 and $\sim$160\,Myr, respectively. These
abundances are 0.4--0.5~dex and 0.1~dex lower than those measured in
the outer layers. These two systems seem to confirm previous evidence
that the properties of Am stars are generally matched by stellar
evolution models just as well as those of normal A stars, suggesting
their properties are not fundamentally different, even though their
surface abundances are. There are exceptions to this rule, though,
such as the recently studied Am system of XY~Cet (Southworth et al.\
\cite{Southworth:11}), whose stars do not seem to agree well with
theory.

Internal structure models predict that apsidal motion in both \SW\ and
\HW\ (empirically determined only in the first case) should be
dominated by General relativistic effects ($\sim$ 50\% and 97\%
contributions, respectively). Current tidal theory suggests that the
angular rotation rates of the \SW\ stars should be pseudo-synchronized
with the orbital motion by now, and that their spin axes should be
parallel to the orbital axis. Indeed we find that the measured
projected rotational velocity of the primary agrees with expectations,
but the measured $v \sin i$ of the secondary is too small by a factor
of two, even though its mass is only 6\% smaller than the
primary. This disagreement cannot be resolved by tuning free
parameters in the models, and underscores our incomplete understanding
of these processes. For \HW\ theory predicts that neither
pseudo-synchronization nor spin-orbit alignment have yet been reached.
However, the measured rotations do agree with the pseudo-synchronous
values in this case, which can nevertheless be explained with a proper
combination of initial values for $\Omega/\Omega_{\rm orb}$ and the
spin-orbit alignment angle $\phi$.


\begin{acknowledgements}

We dedicate this work to our long-time friend and colleague Jens Viggo
Clausen, who passed away as this paper was being prepared for
submission. He was a key participant in this series of articles, and
will be remembered for his high standards, dedication, and careful
attention to detail.  Many of the spectroscopic observations at CfA
were obtained by P.\ Berlind, J.\ Caruso, R.\ Davis, E.\ Horine, A.\
Milone, and J.\ Zajac.  We are grateful to E.\ Sturm for providing his
original disentangling code, and to him and J.\ D.\ Pritchard for
modifying it for use on Linux/Unix computer systems. The
anonymous referee is also thanked for a number of helpful comments.
The projects ``Stellar structure and evolution -- new challenges from
ground and space observations'' and ``Stars: Central engines of the
evolution of the Universe'', carried out at Copenhagen University and
Aarhus University, are supported by the Danish National Science
Research Council.  G.T.\ acknowledges partial support from NSF grant
AST-1007992.  The following internet-based resources were used in the
research for this paper: the NASA Astrophysics Data System, the SIMBAD
database and the VizieR service operated by CDS, Strasbourg, France,
and the ar$\chi$iv scientific paper preprint service operated by
Cornell University.

\end{acknowledgements}


{}

\listofobjects


\clearpage

\begin{appendix}
\section{Radial velocity measurements}
\label{ap:rvtabs}


\begin{table}

\caption[]{\label{tab:swcma_rv} Radial velocities for \SW\ (with all
corrections applied) and residuals from the final spectroscopic
orbit.} 

\begin{center}
\begin{tabular}{lcrrcccc} 
\hline\hline\noalign{\smallskip}

\multicolumn{1}{c}{HJD} &
\multicolumn{1}{c}{Phase} &
\multicolumn{1}{c}{$RV_p$} &
\multicolumn{1}{c}{$RV_s$} &
\multicolumn{1}{c}{$\sigma_p$} &
\multicolumn{1}{c}{$\sigma_s$} &
\multicolumn{1}{c}{$(O-C)_p$} &
\multicolumn{1}{c}{$(O-C)_s$}
\\ 
2\,400\,000+ &
&
\multicolumn{1}{c}{(\kms)} &
\multicolumn{1}{c}{(\kms)} &
\multicolumn{1}{c}{(\kms)} &
\multicolumn{1}{c}{(\kms)} &
\multicolumn{1}{c}{(\kms)} &
\multicolumn{1}{c}{(\kms)} \\

\hline\noalign{\smallskip}
 47254.6732 &  0.1151 &  $-$61.37 &  152.33  &  1.47  &  0.90  &  $-$0.17  &   $-$0.68 \\
 47461.8907 &  0.6480 &  101.18 &  $-$18.34  &  1.90  &  1.17  &   \phs2.75  &   $-$1.45 \\
 47462.8217 &  0.7402 &   97.60 &  $-$16.08  &  1.93  &  1.19  &  $-$1.15  &    \phs1.15 \\
 47463.7887 &  0.8360 &   84.37 &   $-$4.81  &  2.60  &  1.60  &  $-$3.40  &    \phs0.73 \\
 47481.8766 &  0.6283 &   96.92 &  $-$15.02  &  2.00  &  1.23  &  $-$0.35  &    \phs0.63 \\
 47487.7501 &  0.2103 &  $-$36.35 &  130.52  &  2.98  &  1.84  &   \phs2.21  &    \phs1.61 \\
 47487.8417 &  0.2194 &  $-$35.79 &  125.03  &  2.89  &  1.78  &  $-$2.53  &    \phs1.76 \\
 47487.8904 &  0.2242 &  $-$30.05 &  118.29  &  2.70  &  1.66  &   \phs0.34  &   $-$1.93 \\
 47489.7549 &  0.4090 &   52.54 &   26.98  &  2.23  &  1.38  &  $-$5.36  &    \phs0.73 \\
 47489.8541 &  0.4188 &   63.41 &   22.83  &  2.21  &  1.36  &   \phs2.50  &   $-$0.22 \\
 47490.7585 &  0.5084 &   84.03 &    2.51  &  2.12  &  1.31  &   \phs1.74  &    \phs2.21 \\
 47490.8401 &  0.5165 &   78.89 &    0.23  &  2.29  &  1.41  &  $-$4.85  &    \phs1.49 \\
 47491.7770 &  0.6094 &   94.25 &  $-$12.11  &  1.78  &  1.10  &  $-$1.56  &    \phs1.99 \\
 47494.8024 &  0.9091 &   67.85 &   16.06  &  2.38  &  1.47  &   \phs0.61  &   $-$0.25 \\
 47497.7536 &  0.2016 &  $-$42.16 &  134.42  &  1.87  &  1.15  &   \phs1.33  &    \phs0.26 \\
 47497.8143 &  0.2076 &  $-$39.65 &  130.50  &  1.86  &  1.15  &   \phs0.48  &   $-$0.08 \\
 47511.7356 &  0.5870 &   93.81 &  $-$11.36  &  1.83  &  1.13  &   \phs0.15  &    \phs0.45 \\
 47513.8367 &  0.7952 &   91.70 &  $-$13.33  &  2.09  &  1.29  &  $-$2.49  &   $-$0.96 \\
 47514.7963 &  0.8903 &   74.47 &    9.37  &  2.03  &  1.25  &   \phs0.60  &    \phs0.12 \\
 47517.7110 &  0.1791 &  $-$56.28 &  147.56  &  2.38  &  1.47  &  $-$1.66  &    \phs1.55 \\
 47517.7960 &  0.1875 &  $-$52.30 &  144.11  &  2.21  &  1.36  &  $-$1.53  &    \phs2.21 \\
 47518.6764 &  0.2748 &   $-$0.51 &   88.84  &  1.93  &  1.19  &  $-$0.02  &    \phs0.45 \\
 47522.7188 &  0.6753 &   97.25 &  $-$15.26  &  2.48  &  1.53  &  $-$2.19  &    \phs2.70 \\
 47526.6630 &  0.0662 &  $-$35.77 &  124.44  &  1.86  &  1.15  &  $-$1.80  &    \phs0.42 \\
 47527.6974 &  0.1687 &  $-$59.26 &  150.60  &  1.61  &  0.99  &  $-$0.55  &    \phs0.25 \\
 47527.7595 &  0.1748 &  $-$60.16 &  145.03  &  3.77  &  2.32  &  $-$3.75  &   $-$2.88 \\
 47538.7108 &  0.2600 &   $-$9.90 &   96.62  &  1.85  &  1.14  &  $-$0.88  &   $-$0.85 \\
 47540.6888 &  0.4559 &   70.50 &   12.97  &  1.58  &  0.97  &  $-$0.51  &    \phs0.67 \\
 47543.6743 &  0.7518 &   98.52 &  $-$17.52  &  1.47  &  0.91  &   \phs0.39  &   $-$0.95 \\
 47544.6361 &  0.8471 &   81.77 &   $-$2.82  &  1.86  &  1.15  &  $-$3.72  &    \phs0.29 \\
 47546.6844 &  0.0500 &  $-$18.42 &  108.97  &  1.72  &  1.06  &   \phs3.01  &   $-$1.71 \\
 47547.7765 &  0.1583 &  $-$59.76 &  153.21  &  1.53  &  0.94  &   \phs2.05  &   $-$0.44 \\
 47548.6348 &  0.2433 &  $-$18.90 &  107.30  &  1.75  &  1.08  &   \phs0.02  &   $-$0.71 \\
 47550.7033 &  0.4483 &   70.18 &   11.96  &  1.61  &  0.99  &   \phs1.11  &   $-$2.40 \\
 47554.7119 &  0.8455 &   88.11 &   $-$4.06  &  1.91  &  1.18  &   \phs2.28  &   $-$0.58 \\
 47558.6735 &  0.2380 &  $-$22.00 &  112.56  &  1.75  &  1.08  &   \phs0.10  &    \phs1.17 \\
 47577.5485 &  0.1083 &  $-$61.22 &  151.79  &  1.73  &  1.07  &  $-$2.30  &    \phs1.21 \\
 47577.5732 &  0.1108 &  $-$61.14 &  151.22  &  1.59  &  0.98  &  $-$1.33  &   $-$0.31 \\
 47577.6135 &  0.1148 &  $-$60.68 &  152.79  &  1.78  &  1.10  &   \phs0.42  &   $-$0.11 \\
 47587.5326 &  0.0976 &  $-$53.93 &  146.30  &  1.46  &  0.90  &   \phs0.27  &    \phs0.75 \\
 47587.6020 &  0.1045 &  $-$58.35 &  148.81  &  1.68  &  1.04  &  $-$0.96  &   $-$0.14 \\
\noalign{\smallskip}
\hline
\end{tabular}
\end{center}
\end{table}

\clearpage


\begin{table}

\caption[]{\label{tab:hwcma_rv} Radial velocities for \HW\ (with all
corrections applied) and residuals from the final spectroscopic
orbit.} 

\begin{center}
\begin{tabular}{lcrrcccc} 
\hline\hline\noalign{\smallskip}

\multicolumn{1}{c}{HJD} &
\multicolumn{1}{c}{Phase} &
\multicolumn{1}{c}{$RV_s$} &
\multicolumn{1}{c}{$RV_p$} &
\multicolumn{1}{c}{$\sigma_s$} &
\multicolumn{1}{c}{$\sigma_p$} &
\multicolumn{1}{c}{$(O-C)_s$} &
\multicolumn{1}{c}{$(O-C)_p$}
\\ 
2\,400\,000+ &
&
\multicolumn{1}{c}{(\kms)} &
\multicolumn{1}{c}{(\kms)} &
\multicolumn{1}{c}{(\kms)} &
\multicolumn{1}{c}{(\kms)} &
\multicolumn{1}{c}{(\kms)} &
\multicolumn{1}{c}{(\kms)} \\

\hline\noalign{\smallskip}
 47480.7978  &  0.7569 &  $-$26.24 &  60.25 &   0.89  &  0.99  &   $-$0.15  &   $-$1.16 \\
 47480.9295  &  0.7632 &  $-$27.92 &  63.17 &   1.14  &  1.26  &   $-$0.59  &    \phs0.48 \\
 47481.8651  &  0.8075 &  $-$35.51 &  72.50 &   0.91  &  1.01  &    \phs0.78  &    \phs0.54 \\
 47483.8058  &  0.8994 &  $-$50.13 &  86.33 &   0.98  &  1.08  &    \phs1.08  &   $-$1.07 \\
 47513.8485  &  0.3220 &   48.76 & $-$12.14 &   1.61  &  1.78  &    \phs1.46  &    \phs2.39 \\
 47514.8083  &  0.3675 &   40.24 &  $-$4.21 &   1.50  &  1.66  &    \phs0.47  &    \phs2.52 \\
 47518.6906  &  0.5513 &    9.00 &  23.99 &   1.39  &  1.54  &   $-$1.31  &    \phs0.24 \\
 47526.6848  &  0.9298 &  $-$50.92 &  81.32 &   2.02  &  2.25  &   $-$2.09  &   $-$3.61 \\
 47538.7218  &  0.4998 &   15.50 &  16.45 &   1.27  &  1.41  &   $-$3.06  &    \phs1.24 \\
 47540.7012  &  0.5936 &    3.93 &  29.51 &   0.94  &  1.04  &    \phs0.55  &   $-$1.41 \\
 47543.6839  &  0.7348 &  $-$22.42 &  56.76 &   0.77  &  0.86  &   $-$0.66  &   $-$0.17 \\
 47544.6634  &  0.7812 &  $-$31.11 &  65.89 &   0.91  &  1.01  &   $-$0.17  &   $-$0.54 \\
 47546.7040  &  0.8778 &  $-$48.03 &  84.78 &   0.91  &  1.01  &    \phs1.18  &   $-$0.55 \\
 47548.6480  &  0.9699 &  $-$24.31 &  59.46 &   0.87  &  0.96  &    \phs0.25  &   $-$0.37 \\
 47550.7274  &  0.0683 &   78.92 & $-$44.53 &   1.08  &  1.20  &    \phs2.00  &    \phs0.64 \\
 47554.7299  &  0.2579 &   57.28 & $-$27.08 &   0.95  &  1.06  &   $-$1.05  &   $-$1.15 \\
 47558.6948  &  0.4456 &   25.54 &   8.17 &   1.39  &  1.54  &   $-$1.65  &    \phs1.88 \\
 47568.6065  &  0.9150 &  $-$52.42 &  86.64 &   0.96  &  1.06  &   $-$1.36  &   $-$0.60 \\
 47569.5913  &  0.9616 &  $-$32.87 &  69.07 &   0.89  &  0.99  &   $-$0.49  &    \phs1.15 \\
 47569.6698  &  0.9653 &  $-$27.99 &  63.11 &   0.91  &  1.00  &    \phs1.08  &   $-$1.38 \\
 47570.6072  &  0.0097 &   28.93 &   7.99 &   0.84  &  0.94  &    \phs0.29  &    \phs3.20 \\
 47577.5864  &  0.3402 &   45.15 & $-$13.62 &   0.83  &  0.93  &    \phs0.89  &   $-$2.24 \\
 47864.8542  &  0.9433 &  $-$45.75 &  80.66 &   1.45  &  1.61  &   $-$1.54  &    \phs0.50 \\
 47865.7571  &  0.9861 &   $-$7.69 &  40.86 &   1.87  &  2.07  &   $-$2.72  &    \phs1.30 \\
 47866.7874  &  0.0348 &   59.01 & $-$25.42 &   1.30  &  1.45  &    \phs1.15  &    \phs0.03 \\
 47868.7776  &  0.1291 &   79.34 & $-$49.86 &   1.32  &  1.46  &    \phs0.01  &   $-$2.20 \\
 47869.8006  &  0.1775 &   73.15 & $-$41.34 &   1.18  &  1.31  &    \phs0.77  &   $-$0.87 \\
 47883.7709  &  0.8391 &  $-$42.56 &  78.95 &   0.89  &  0.99  &    \phs0.05  &    \phs0.45 \\
 47903.7315  &  0.7843 &  $-$31.93 &  66.06 &   0.93  &  1.04  &   $-$0.36  &   $-$1.01 \\
 47904.7120  &  0.8307 &  $-$40.79 &  77.37 &   0.91  &  1.00  &    \phs0.18  &    \phs0.56 \\
 47905.6679  &  0.8760 &  $-$49.10 &  85.56 &   0.93  &  1.03  &   $-$0.14  &    \phs0.49 \\
 47906.7301  &  0.9263 &  $-$49.20 &  84.82 &   1.52  &  1.69  &    \phs0.40  &   $-$0.91 \\
 47930.6134  &  0.0572 &   71.99 & $-$41.26 &   0.95  &  1.06  &   $-$0.82  &   $-$0.34 \\
 47934.6052  &  0.2462 &   58.54 & $-$27.98 &   1.70  &  1.89  &   $-$1.83  &    \phs0.07 \\
 47942.5941  &  0.6245 &   $-$0.84 &  36.36 &   0.87  &  0.97  &    \phs0.98  &    \phs0.06 \\
 48727.6172  &  0.7980 &  $-$33.23 &  70.24 &   0.74  &  0.82  &    \phs1.13  &    \phs0.27 \\
 49316.8629  &  0.7008 &  $-$16.18 &  50.58 &   1.03  &  1.14  &   $-$0.85  &    \phs0.30 \\
 49330.7929  &  0.3604 &   40.55 &  $-$7.91 &   1.12  &  1.24  &   $-$0.37  &    \phs0.02 \\
 49400.6145  &  0.6667 &   $-$9.17 &  44.14 &   0.89  &  0.99  &   $-$0.02  &    \phs0.26 \\
 50003.0181  &  0.1925 &   70.08 & $-$37.97 &   0.48  &  0.53  &    \phs0.25  &   $-$0.14 \\
 50027.8831  &  0.3699 &   37.97 &  $-$6.92 &   0.96  &  1.06  &   $-$1.39  &   $-$0.61 \\
 50030.9707  &  0.5162 &   17.50 &  17.33 &   0.86  &  0.95  &    \phs1.54  &   $-$0.57 \\
 50033.9660  &  0.6580 &   $-$8.13 &  41.78 &   0.82  &  0.91  &   $-$0.52  &   $-$0.51 \\
 50034.9653  &  0.7053 &  $-$16.29 &  51.75 &   0.61  &  0.68  &   $-$0.12  &    \phs0.60 \\
 50059.8947  &  0.8858 &  $-$50.61 &  86.64 &   0.55  &  0.61  &   $-$0.44  &    \phs0.32 \\
 50088.8617  &  0.2575 &   58.02 & $-$25.57 &   0.50  &  0.55  &   $-$0.38  &    \phs0.43 \\
 50092.8003  &  0.4440 &   27.65 &   5.77 &   0.88  &  0.98  &    \phs0.20  &   $-$0.25 \\
 51239.7317  &  0.7550 &  $-$24.38 &  61.40 &   0.87  &  0.97  &    \phs1.33  &    \phs0.38 \\
\noalign{\smallskip}
\hline
\end{tabular}
\end{center}
\end{table}

\clearpage


\section{Light curve solutions for \SW\ under different limb-darkening assumptions}
\label{ap:LDswcma}

Tables~\ref{tab:swcma_ebop_vh} and \ref{tab:swcma_ebop_claret} present
our JKTEBOP solutions for fixed linear LD coefficients ($u_p$, $u_s$)
from van Hamme (\cite{vh93}) and from Claret (\cite{Claret:00}),
respectively. The uncertainties we report are formal errors from the
iterative least-squares procedure, which are often found to be
underestimated. In each case the agreement between the $uvby$
passbands is quite good, and the results using the different LD
coefficients show small but perhaps significant systematic
differences, with the inclination angle being about $0{\fdg}22$
smaller using the Claret values, the sum of the radii being
$\sim$0.5\% larger, and $k$ also being 1--3\% larger (resulting in
$r_p$ values up to 1\% smaller, and $r_s$ values 1--3\%
larger). Experiments using Claret coefficients for a quadratic law
instead of the linear law lead to solutions that are not significantly
better and results that are not very different, and tend to be closer
to those obtained with the van Hamme coefficients.

\begin{table}[b]

\caption{Photometric solutions for \SW\ with linear limb darkening
coefficients adopted from van Hamme (\cite{vh93}).}

\label{tab:swcma_ebop_vh}
\begin{center}
\begin{tabular}{lrrrr} 
\hline\hline
\noalign{\smallskip}
  Parameter          &     $y$~~    &       $b$~~  &       $v$~~  &   $u$~~~\\                   
\noalign{\smallskip}
\hline
\noalign{\smallskip}
$i$  (\degr)         &  88.58     &   88.71    &   88.75    &  88.53\vspace{-0.8mm}\\   
                     & $\pm 5$    &  $\pm 3$   &  $\pm 3$   & $\pm 7$\\                 

$e\cos \omega$       &$-0.30399$  & $-0.30314$ & $-0.30318$ &$-0.30318$\vspace{-0.8mm}\\ 
                     &  $\pm  5$  &   $\pm  5$ &   $\pm  5$ &  $\pm  8$\\               

$e\sin \omega$       &$+0.0906$   & $+0.0880$  & $+0.0877$  &$+0.0900$\vspace{-0.8mm}\\  
                     & $\pm 13$   &  $\pm 12$  & $\pm  12$  & $\pm 20$\\                

$e$                  &  0.3163    &   0.3157   &   0.3156   &  0.3163\\                 
                                                                                        
$\omega$  (\degr)    & 163.35     &  163.81    &  163.87    & 163.46 \\                 
                                                                                        
$r_p + r_s$          &  0.1718    &   0.1714   &   0.1712   &  0.1721\vspace{-0.8mm}\\  
                     &  $\pm 3$   &   $\pm 3$  &   $\pm 3$  &  $\pm 4$ \\               
                                                                                        
$k$                  &   0.829    &   0.820    &   0.819    &   0.834\vspace{-0.8mm}\\  
                     &  $\pm 7$   &   $\pm4$   &   $\pm3$   &  $\pm11$ \\               

$r_p$                &  0.0939    &   0.0942   &   0.0941   &  0.0938\\                 

$r_s$                &  0.0779    &   0.0772   &   0.0771   &  0.0782\\                 
                                                                                        
$u_p$                &  0.50      &   0.59     &   0.63     &  0.52\\      

$u_s$                &  0.51      &   0.59     &   0.62     &  0.54\\      


$y_p$                &  0.83      &   0.95     &   1.07     &  1.26\\

$y_s$                &  0.84      &   0.96     &   1.08     &  1.28\\

$J_s$                &  1.0153    &    1.0088  &   0.9884   &  1.0269\vspace{-0.8mm}\\
                     &  $\pm13$   &   $\pm 12$ &   $\pm12$  &  $\pm19$\\

$L_s/L_p$            &  0.6946    &    0.6777  &   0.6652   &  0.7075 \\

$\sigma$  (mmag)  &  5.8       &    5.4     &   5.5      &  8.0   \\

$N_{\rm obs}$ used   &  818       &    815     &   820     &   820 \\
\noalign{\smallskip}\hline
\end{tabular}            
\end{center}            
\end{table}                       


\begin{table}[t]

\caption{Photometric solutions for \SW\ with linear limb darkening
coefficients adopted from Claret (\cite{Claret:00}).}

\label{tab:swcma_ebop_claret}
\begin{center}
\begin{tabular}{lrrrr} 
\hline\hline
\noalign{\smallskip}
    Parameter      &     $y$~~    &       $b$~~  &       $v$~~  &   $u$~~~\\                   
\noalign{\smallskip}\hline\noalign{\smallskip}            
$i$  (\degr)       &  88.34     &   88.47    &   88.54    &  88.32\vspace{-0.8mm}\\   
                     & $\pm 5$    &  $\pm 4$   &  $\pm 4$   & $\pm 7$\\                 

$e\cos \omega$       &$-0.30295$  & $-0.30312$ & $-0.30317$ &$-0.30311$\vspace{-0.8mm}\\ 
                     &  $\pm  7$  &   $\pm  5$ &   $\pm  5$ &  $\pm  9$\\               

$e\sin \omega$       &$+0.0911$   & $+0.0878$  & $+0.0870$  &$+0.0915$\vspace{-0.8mm}\\  
                     & $\pm 18$   &  $\pm 15$  & $\pm  15$  & $\pm 26$\\                

$e$                  &  0.3163    &   0.3156   &   0.3154   &  0.3166\\                 
                                                                                   
$\omega$  (\degr)  & 163.27     &  163.84    &  164.00    & 163.20 \\                 

$r_p + r_s$          &  0.1729    &   0.1725   &   0.1721   &  0.1730\vspace{-0.8mm}\\  
                     &  $\pm 3$   &   $\pm 3$  &   $\pm 6$  &  $\pm 5$ \\               
$k$                  &   0.855    &   0.836    &   0.830    &   0.864\vspace{-0.8mm}\\  
                     &  $\pm12$   &   $\pm7$   &   $\pm6$   &  $\pm19$ \\               

$r_p$                &  0.0932    &   0.0939   &   0.0940   &  0.0929\\                 

$r_s$                &  0.0797    &   0.0786   &   0.0781   &  0.0802\\                 
                                                                                       
$u_p$                &  0.60      &   0.69     &   0.72     &  0.59\\                 

$u_s$                &  0.61      &   0.69     &   0.72     &  0.60\\                

$y_p$                &  0.83      &   0.95     &   1.07     &  1.26\\

$y_s$                &  0.84      &   0.96     &   1.08     &  1.28\\

$J_s$                &  1.0207    &    1.0142  &   0.9977   &  1.0255\vspace{-0.8mm}\\
                     &  $\pm18$   &   $\pm 18$ &   $\pm18$  &  $\pm26$\\

$L_s/L_p$            &  0.7430    &    0.7085  &   0.6871   &  0.7606 \\

$\sigma$  (mmag)  &  5.8       &    5.5     &   5.5      &  7.9   \\

$N_{\rm obs}$ used   &  818       &    817     &   820     &   820 \\

\noalign{\smallskip}\hline
\end{tabular}            
\end{center}            
\end{table}                       


\begin{table}[b!]
\begin{center}

\caption{Photometric solutions for \SW\ allowing the linear limb
darkening coefficients to vary freely, with the condition that they be
the same for the two components.}

\label{tab:swcma_ebop_free}
\begin{tabular}{lrrrr} 
\hline\hline
\noalign{\smallskip}
  Parameter        &     $y$~~    &       $b$~~  &       $v$~~  &   $u$~~~\\                   
\noalign{\smallskip}\hline\noalign{\smallskip}
$i$  (\degr)       &  88.53     &   88.59    &   88.62    &  88.62\vspace{-0.8mm}\\   
                     & $\pm16$    &  $\pm12$   &  $\pm11$   & $\pm16$\\                 

$e\cos \omega$       &$-0.30300$  & $-0.30314$ & $-0.30318$ &$-0.30321$\vspace{-0.8mm}\\ 
                     &  $\pm  6$  &   $\pm  5$ &   $\pm  5$ &  $\pm  7$\\               

$e\sin \omega$       &$+0.0902$   & $+0.0875$  & $+0.0869$  &$+0.0893$\vspace{-0.8mm}\\  
                     & $\pm 15$   &  $\pm 14$  & $\pm  14$  & $\pm 18$\\                

$e$                  &  0.3161    &   0.3155   &   0.3154   &  0.3161\\                 
                                                                                        
$\omega$  (\degr)  & 163.45     &  163.89    &  164.00    & 163.59 \\                 

$r_p + r_s$          &  0.1720    &   0.1719   &   0.1717   &  0.1716\vspace{-0.8mm}\\  
                     &  $\pm 8$   &   $\pm 6$  &   $\pm 6$  &  $\pm 8$\\                
                                                                                        
$k$                  & 0.832      &   0.826    &   0.825    &  0.824\vspace{-0.8mm}\\
                     & $\pm 18$   &  $\pm 10$  &  $\pm 8$   & $\pm 13$\\

$r_p$                &  0.0939    &   0.0941   &   0.0941   &  0.0941\\  

$r_s$                &  0.0781    &   0.0778   &   0.0776   &  0.0775\\                 
                                                                                        
$u_p = u_s$          &  0.52      &   0.64     &   0.69     &  0.50\vspace{-0.8mm}\\ 
                     & $\pm5$     &  $\pm4$    &  $\pm4$    & $\pm6$             \\

$y_p$                &  0.83      &   0.95     &   1.07     &  1.26\\

$y_s$                &  0.84      &   0.96     &   1.08     &  1.28\\

$J_s$                &  1.0136    &    1.0121  &   0.9961   &  1.0188\vspace{-0.8mm}\\
                     &  $\pm36$   &   $\pm 26$ &   $\pm24$  &  $\pm50$\\

$L_s/L_p$            &  0.7013    &    0.6903  &   0.6771   &  0.6903 \\

$\sigma$  (mmag)  &  5.8       &    5.5     &   5.5      &  8.0   \\

$N_{\rm obs}$ used   &  818       &    817     &   820     &   820 \\
\noalign{\smallskip}\hline
\end{tabular}                       
\end{center}
\end{table}            


Solutions with the LD coefficients free are presented in
Table~\ref{tab:swcma_ebop_free}, subject only to the condition that
the coefficients be the same for the primary and secondary since their
temperatures are also very similar. These results indicate slightly
better agreement than before between the four passbands. On average
the fitted LD coefficients are closer to those by van Hamme than those
by Claret. We summarize the elements obtained from the three different
LD prescriptions in Table~\ref{tab:swcma_ebop_mean}, where the results
from the separate passbands have been averaged in each case, with
weights inversely proportional to the rms residual of each
solution. The light elements finally adopted for the analysis of \SW\
are those with LD free, and are repeated in Table~\ref{tab:swcma_phel}
of Sect.~\ref{sec:swcma_phel} in the main text, with more conservative
errors as described there.

\begin{table}
\begin{center}

\caption{Weighted mean photometric solutions for \SW\ for three
different treatments of the linear limb darkening coefficients.}

\label{tab:swcma_ebop_mean}
\begin{tabular}{lrrr} 
\hline\hline\noalign{\smallskip}
 Parameter         &  Free      &  van Hamme &  Claret    \\
\noalign{\smallskip}\hline\noalign{\smallskip}
$i$  (\degr)       &  88.59     &   88.66    &   88.44\vspace{-0.8mm}\\
                     & $\pm 4$    &  $\pm10$   &  $\pm10$   \\

$e\cos \omega$       &$-0.30313$  & $-0.30312$ & $-0.30309$\vspace{-0.8mm}\\
                     &  $\pm  9$  &   $\pm  9$ &   $\pm 10$ \\

$e\sin \omega$       &$+0.0883$   & $+0.0889$  & $+0.0889$\vspace{-0.8mm}\\
                     & $\pm 16$   &  $\pm 15$  & $\pm  22$  \\

$e$                  &  0.3157    &   0.3159   &   0.3159\vspace{-0.8mm}\\
                     & $\pm  4$   &   $\pm  4$  &  $\pm  6$   \\
                                                              
$\omega$  (\degr)  & 163.76     &  163.66    &  163.65\vspace{-0.8mm}\\ 
                     &$\pm 28$    & $\pm 26$   & $\pm 39$     \\
                                                               
$r_p + r_s$          &  0.1718    &   0.1715   &   0.1726\vspace{-0.8mm}\\
                     & $\pm  1$   &  $\pm  3$  &  $\pm  4$   \\

$k$                  &   0.827    &   0.824    &   0.843\vspace{-0.8mm}\\
                     &  $\pm 4$   &  $\pm 7$   &  $\pm15$  \\

$r_p$                &  0.0940    &   0.0940   &   0.0936\vspace{-0.8mm}\\
                     &  $\pm 1$   &   $\pm 2$  &   $\pm 5$  \\

$r_s$                &  0.0778    &   0.0775   &   0.0789\vspace{-0.8mm}\\
                     & $\pm  2$   &  $\pm  5$  &  $\pm  9$   \\
                                                             
\noalign{\smallskip}\hline
\end{tabular}            
\end{center}
\end{table}                       


\section{Light curve solutions for \HW\ under different limb-darkening assumptions}
\label{ap:LDhwcma}

Table~\ref{tab:hwcma_ebop_claret} reports the results from JKTEBOP
fits using linear LD coefficients from Claret (\cite{Claret:00}).  The
errors listed include the uncertainty in the light ratio constraint,
but are otherwise internal and unrealistically small in most
cases. Similarly small errors are obtained using the van Hamme
(\cite{vh93}) coefficients. The geometric elements show good agreement
between the $v$, $b$, and $y$ bands, with $u$ being more discrepant
(and also more uncertain).  Results using the LD coefficients from van
Hamme (\cite{vh93}) are given in Table~\ref{tab:hwcma_ebop_vh}, and
show similarly good agreement in $vby$.  In this case we report more
conservative errors from 1000 Monte Carlo simulations in which we
perturbed the main adjustable quantities that were held fixed to allow
their errors to propagate through: the theoretical LD coefficients
were allowed to vary by $\pm 0.08$, $e \sin\omega$ and $e \cos\omega$
were perturbed by amounts corresponding to the spectroscopic
uncertainties in $e$ and $\omega$, and the flux ratios $J_s$ were
allowed to vary by $\pm 0.002$. Monte Carlo errors using LD from
Claret (\cite{Claret:00}) are very similar to these.

The systematic differences between the fits with Claret and van Hamme
LD coefficients are smaller than we found before for \SW: with the
Claret coefficients the inclination angle is marginally smaller (by
0\fdg01), the sum of the radii is $\sim$0.5\% larger, and $k$ is also
0.1\% larger, all in the same direction as found for \SW. The
individual radii are both systematically larger by about 0.5\%.

The light elements we adopt for the analysis of \HW\ are those that
use the van Hamme (\cite{vh93}) coefficients. Final values averaged
over the $vby$ passbands are presented in Table~\ref{tab:hwcma_phel}
of Sect.~\ref{sec:hwcma_phel}.

\begin{table}

\caption{Constrained photometric solutions for \HW\ with linear limb
darkening coefficients adopted from Claret (\cite{Claret:00}). 
}

\label{tab:hwcma_ebop_claret}
\begin{center}
\begin{tabular}{lrrrr} 
\hline\hline
\noalign{\smallskip}
   Parameter         &     $y$~~    &       $b$~~  &       $v$~~  &   $u$~~~\\                   
\noalign{\smallskip}
\hline
\noalign{\smallskip}
$i$  (\degr)         &  84.79     &   84.85    &   84.85    &  84.66\\   
                     &  $\pm 6$   &   $\pm 7$  &  $\pm 5$    &  $\pm 9$\\

$r_p + r_s$          &  0.06862    &   0.06763   &   0.06781   &  0.07038\\  
                     &  $\pm 6$    &  $\pm 8$    &   $\pm 6$   &  $\pm 10$\\

$k$                  &  1.01294    &   1.01275    &   1.01304    &   1.01076\\  
                     &  $\pm 4$    &  $\pm 5$    &   $\pm 4$   &  $\pm 5$\\

$r_p$                &  0.03409    &   0.03360   &   0.03368   &  0.03500\\                 

$r_s$                &  0.03453   &   0.03403   &   0.03412   &  0.03538\\                 
                                                                                        
$u_p$                &  0.59      &   0.67     &   0.70     &  0.65\\      

$u_s$                &  0.59      &   0.67     &   0.70     &  0.64\\      

$y_p$                &  0.90      &   1.03     &   1.17     &  1.38\\

$y_s$                &  0.88      &   1.01     &   1.15     &  1.36\\

$J_s$                &  1.075     &    1.091   &   1.105   &  1.078\\

$L_s/L_p$            &  1.103     &    1.119  &   1.134   &  1.106 \\

$\sigma$  (mmag)     &  4.4       &    5.5     &   4.1      &  7.0   \\

$N_{\rm obs}$ used   &  415       &    413     &   409     &   413 \\
\noalign{\smallskip}\hline
\end{tabular}            
\end{center}            
\end{table}                       


\begin{table}

\caption{Constrained photometric solutions for \HW\ with linear limb
darkening coefficients adopted from van Hamme (\cite{vh93}).}
\label{tab:hwcma_ebop_vh}
\begin{center}
\begin{tabular}{lrrrr} 
\hline\hline
\noalign{\smallskip}
   Parameter         &     $y$~~    &       $b$~~  &       $v$~~  &   $u$~~~\\                   
\noalign{\smallskip}
\hline
\noalign{\smallskip}
$i$  (\degr)         &  84.80     &   84.87    &   84.87    &  84.67\vspace{-0.8mm}\\   
                     &  $\pm 8$   &   $\pm 9$  &  $\pm 8$    &  $\pm 11$\\

$r_p + r_s$          &  0.06829    &   0.06725   &   0.06741   &  0.07008\vspace{-0.8mm}\\  
                     &  $\pm 67$    &  $\pm 81$    &   $\pm 66$   &  $\pm 109$\\

$k$                  &  1.0129    &   1.0107    &   1.0109    &   1.0108\vspace{-0.8mm}\\  
                     &  $\pm 225$    &  $\pm 223$    &   $\pm 223$   &  $\pm 227$\\

$r_p$                &  0.03393    &   0.03345   &   0.03352   &  0.03485\vspace{-0.8mm}\\
                     &  $\pm 45$    &  $\pm 50$    &   $\pm 49$   &  $\pm 64$\\

$r_s$                &  0.03437   &   0.03380   &   0.03389   &  0.03523\vspace{-0.8mm}\\
                     &  $\pm 55$    &  $\pm 57$    &   $\pm 53$   &  $\pm 69$\\

$u_p$                &  0.51      &   0.59     &   0.62     &  0.58\\      

$u_s$                &  0.51      &   0.58     &   0.61     &  0.57\\      

$y_p$                &  0.90      &   1.03     &   1.17     &  1.38\\

$y_s$                &  0.88      &   1.01     &   1.15     &  1.36\\

$J_s$                &  1.075     &    1.091   &   1.105   &  1.078\vspace{-0.8mm}\\
                     &  $\pm 2$    &  $\pm 2$    &   $\pm 2$   &  $\pm 2$\\

$L_s/L_p$            &  1.103     &    1.119  &   1.134   &  1.106\vspace{-0.8mm} \\
                     &  $\pm 20$    &  $\pm 20$    &   $\pm 21$   &  $\pm 20$\\

$\sigma$  (mmag)     &  4.4       &    5.5     &   4.1      &  7.0   \\

$N_{\rm obs}$ used   &  415       &    413     &   409     &   413 \\
\noalign{\smallskip}\hline
\end{tabular}            
\end{center}            
\end{table}                       


\end{appendix}


\end{document}